\documentclass{article}

\usepackage[utf8]{inputenc}
\usepackage{amssymb}
\usepackage{amsmath}
\usepackage{natbib}
\usepackage{graphicx}
\usepackage{multirow}
\usepackage{bbm}
\usepackage{mdwlist}
\usepackage{hyperref}
\usepackage{flushend}
\usepackage{microtype}
\usepackage{subfigure}
\usepackage{booktabs}

\usepackage{todonotes}

\newtheorem{theorem}{Theorem}
\newtheorem{definition}{Definition}
\newtheorem{example}{Example}
\newtheorem{corollary}{Corollary}
\newtheorem{lemma}{Lemma}

\usepackage{algorithm}
\usepackage{algorithmic}
\begin{document}

% \twocolumn[
% \icmltitle{Rate Distortion For Model Compression:\\ From Theory To Practice}

% \begin{icmlauthorlist}
% \icmlauthor{Weihao Gao}{uiuc}
% \icmlauthor{Yu-Han Liu}{google}
% \icmlauthor{Chong Wang}{google}
% \icmlauthor{Sewoong Oh}{uw}
% \end{icmlauthorlist}

% \icmlaffiliation{uiuc}{Department of Electrical and Computer Engineering, University of Illinois at Urbana-Champaign. Work done as an intern in Google Inc.}
% \icmlaffiliation{google}{Google, Inc.}
% \icmlaffiliation{uw}{Department of Computer Science, University of Washington.}

% \icmlcorrespondingauthor{Weihao Gao}{wgao9@illinois.edu}
% \vskip 0.3in
% ]

% \printAffiliationsAndNotice{}

\title{Rate Distortion For Model Compression: From Theory To Practice}

\author{
Weihao Gao\thanks{Department of Electrical and Computer Engineering, University of Illinois at Urbana-Champaign. Email: \texttt{wgao9@illinois.edu}, work done as an intern in Google Inc.}, \;\;  
Yu-Han Liu\thanks{Google Inc. Email: \texttt{\{yuhanliu, chongw\}@google.com}},\;\;
Chong Wang$^{\dagger}$, \;\;
Sewoong Oh\thanks{Allen School of Computer Science and Engineering. Email: \texttt{sewoong@cs.washington.edu}}
}

\date{\today}

\maketitle

\begin{abstract}
The enormous size of modern deep neural networks makes it challenging to deploy those models in 
memory and communication limited scenarios. 
%On the other hand, those models are commonly required in small devices with limited storage and bandwidth. 
Thus, compressing a trained model without a significant loss in performance has become an increasingly important task. Tremendous advances has been made recently, where the main technical building blocks are parameter pruning, parameter sharing  (quantization), and low-rank factorization. 
In this paper, we propose principled approaches to improve upon the common heuristics used in those building blocks, namely pruning and quantization.

We first study the fundamental limit for model compression via the rate distortion theory. We bring the rate distortion function from data compression to model compression to quantify this fundamental limit. We prove a lower bound for the rate distortion function and prove its achievability for linear models.  Although this achievable compression scheme is intractable in practice, this analysis motivates a novel model compression framework.
This framework provides a new objective function in 
model compression, which can be applied together with other classes of model compressor such as pruning or quantization. 
Theoretically, we prove that the proposed scheme is optimal for compressing one-hidden-layer ReLU neural networks. Empirically, we show that the proposed scheme improves upon the baseline in the compression-accuracy tradeoff.

%As the size of neural network models increases dramatically today, study of model compression algorithms becomes important. Despite many practically successful compression methods, the fundamental limit of model compression remains unknown. In this paper, we study the fundamental limit for model compression via rate distortion theory. We bring the rate distortion function from data compression to model compression to quantify the fundamental limit. We prove a lower bound for the rate distortion function and prove its achievability for linear models. Motivated by our theory, we further present an improved objective for compression algorithms, and show that the proposed objective is optimal for one-hidden-layer ReLU neural networks. We then demonstrate its good performance for real neural network models.

\end{abstract}

\section{Introduction}

Deep neural networks have been successful, for example, in the application of computer vision~\citep{krizhevsky2012imagenet}, machine translation~\citep{wu2016google} and game playing~\citep{silver2017mastering}. 
With increasing data and computational power, 
%While computing resources become more accessible and neural networks become more powerful, 
the number of weights in practical neural network model also grows rapidly. For example, in the application of image recognition, the LeNet-5 model~\citep{lecun1998gradient} only has 400K weights. After two decades,  AlexNet~\citep{krizhevsky2012imagenet} has more than 60M weights, and VGG-16 net~\citep{simonyan2014very} has more than 130M weights. ~\citet{coates2013deep} even tried a neural network with 11B weights. The huge size of neural networks brings many challenges, including large storage, difficulty in training, and large energy consumption. In particular, 
deploying such extreme models to embedded mobile systems is not feasible. 

Several approaches have been proposed to reduce the size of large neural networks while preserving the performance as much as possible. Most of those  approaches fall into one of the two broad categories. The first category designs novel network structures with small number of parameters, such as SqueezeNet~\cite{iandola2016squeezenet} and MobileNet~\cite{howard2017mobilenets}. The other category directly compresses a given large neural network using pruning, quantization, and matrix factorization, including~\cite{lecun1990optimal,hassibi1993second,han2015learning,han2015deep,cheng2015exploration}. There are also advanced methods to train the neural network using Bayesian methods to help pruning or quantization at a later stage,  such as~\cite{ullrich2017soft,louizos2017bayesian,federici2017improved}.

As more and more model compression algorithms are proposed and compression ratio becomes larger and larger, it motivates us to think about the fundamental question --- How well can we do for model compression? The goal of model compression is to trade off the {\em number of bits} used to describe the model parameters, and the {\em distortion} between the compressed model and original model. We wonder {\em at least} how many bits is needed to achieve certain distortion? Despite many successful model compression algorithms, these theoretical questions still remain unclear.

In this paper, we fill in this gap by bringing tools from rate distortion theory to identify the fundamental limit on how much a model can be compressed. 
Specifically, we focus on compression of a pretrained model, rather than designing new structures or retraining models. Our approach builds upon rate-distortion theory introduced by~\citet{shannon1959coding} and further developed by~\citet{berger1971rate}. The approach also connects to modeling neural networks as random variables in~\citet{mandt2017stochastic}, which has many practical usages~\citep{cao2018review}.
%Such modeling neural networks has been successful in theoretical understanding of generalization~\citep{russo2015much,xu2017information},

Our contribution for model compression is twofold: theoretical and practical. We first apply theoretical tools from rate distortion theory to provide a lower bound on the fundamental trade-off between {\em rate} (number of bits to describe the model) and {\em distortion} between compressed and original models, and prove the tightness of the lower bound for a linear model. This analysis seamlessly incorporate the structure of the neural network architecture into model compression via backpropagation. Motivated by the theory, we design an improved objective for compression algorithms and show that the improved objective gives optimal pruning and quantization algorithm for one-hidden-layer ReLU neural network, and has better performance in real neural networks as well.

% Figure~\ref{fig:linear_pruning} gives an overview of one of the main results of this paper.  
% %Black line is the lower bound of rate distortion function in~\eqref{eq:linear_lower_bound}. Red line and blue line denotes the rate distortion curve of the proposed pruning algorithm in Section~\ref{sec:pruning} and baseline pruning algorithm. 
% Our proposed algorithm is closer to the fundamental limit than the baseline for linear models. See Section~\ref{sec:pruning}.3 for details of this experiment.
% %\todo{Added explanation}
% \begin{figure}[t]
%     \centering
%     \includegraphics[width=0.25\textwidth]{other_figures/linear_pruning.png}
%     \put(-140,-10){Distortion}
%     \put(-240,80){\rotatebox{90}{Rate}}
%     \caption{Illustration of main results on a linear regression model $f_w(x) = \langle w, x \rangle$. The $x$-axis of figure~\ref{fig:linear_pruning} corresponds to the distortion between original model and compressed model, formally defined in Eq.~\eqref{eq:distortion}, whereas the $y$-axis corresponds to the number of bits to describe the compressed weights. Black solid line: Lower bound of rate distortion function, described in Section~\ref{sec:achieve}. Blue dotted line: Rate distortion curve for the baseline pruning algorithm. Red dot-dashed line: Rate distortion curve off for proposed algorithm in Section~\ref{sec:pruning}. }
%     \label{fig:linear_pruning}
% \end{figure}

The rest of the paper is organized as follows.
\begin{itemize}
    \item In Section~\ref{sec:related}, we briefly review some previous work on model compression.
    \item In Section~\ref{sec:rate_distortion}, we introduce the background of the rate distortion theory for data compression, and formally state the rate distortion theory for model compression. %via Minimum Description Length (MDL) principle.
    \item In Section~\ref{sec:lower_bound}, we give a lower bound of the rate distortion function, which quantifies the fundamental limit for model compression. We then prove that the lower bound is achievable for linear model.
    \item In Section~\ref{sec:relu}, motivated by the achievable compressor for linear model, we proposed an improved objective for model compression, which takes consideration of the sturcture of the neural network. We then prove that the improved objective gives optimal compressor for one-hidden-layer ReLU neural network.
    \item In Section~\ref{sec:experiment}, we demonstrate the empirical performance of the proposed objective on fully-connected neural networks on MNIST dataset and convolutional networks on CIFAR dataset.
\end{itemize}

% ----------------------------------------- 

% Outline:
% \begin{itemize}
% 	\item 1. Introduction
%     \begin{itemize}
%     	\item model compression is important in lightweight neural network
%         \item several heuristics but the fundamental limit is not known
%         \item 
%         \item 
%         \item such modeling neural networks has been successful in theoretical understanding of generalization , and practical guideline for more precise predictions with uncertainty .
%         \item 
%         \item 1. identify the limit, and gap 
%         \item 2. achievable schemes from RD theory provides some guidelines to improve
%         \item 
%         \item REALTED WORK 
%         \item concrete example and latest actual neural networks
%         \item measure of compression 
%         \item measure of loss
%         \item 
%     \end{itemize}
%     \item 2. Rate-distortion lower bound
%     \begin{itemize}
%     	\item Modelling assumption
%         \item Rate distortion theory
%         \item Linear case
%     \end{itemize}
%     \item 3. Efficient Model Compression
%     \begin{itemize}
%     	\item pruning
%         \item re-training
%         \item quantization
%     \end{itemize}
% \end{itemize}

% ----------------------------------------- 
\section{Related work on model compression}
\label{sec:related}
The study of model compression of neural networks appeared as long as neural network was invented. Here we mainly discuss the literature on directly compressing large models, which are more relevant to our work. They usually contain three types of methods --- pruning, quantization and matrix factorization.

Pruning methods set unimportant weights to zero to reduce the number of parameters. Early works of model pruning includes biased weight decay~\citep{hanson1989comparing}, optimal brain damage~\citep{lecun1990optimal} and optimal brain surgeon \citep{hassibi1993second}. Early methods utilize the Hessian matrix of the loss function to prune the weights, however, Hessian matrix is inefficient to compute for modern large neural networks with millions of parameters. More recently, ~\citet{han2015learning} proposed an iterative pruning and retraining algorithm that works for large neural networks.

Quantization, or weight sharing methods group the weights into clusters and use one value to represent the weights in the same group. This category includes fixed-point quantization by~\citet{vanhoucke2011improving}, vector quantization by~\citet{gong2014compressing}, HashedNets by~\citet{chen2015compressing}, Hessian-weighted quantizaiton by~\citet{choi2016towards}.

Matrix factorization assumes the weight matrix in each layer could be factored as a low rank matrix plus a sparse matrix. Hence, storing low rank and sparse matrices is cheaper than storing the whole matrix. This category includes~\citet{denton2014exploiting} and~\citet{cheng2015exploration}.

There are some recent advanced method beyond pruning, quantization and matrix factorization.~\citet{han2015deep} assembles pruning, quantization and Huffman coding to achieve better compression rate. Bayesian methods ~\cite{ullrich2017soft,louizos2017bayesian,federici2017improved} are also used to retrain the model such that the model has more space to be compressed.~\citet{he2018amc} uses reinforcement learning to design a compression algorithm.

Despite these aforementioned works for model compression, no one has studied the fundamental limit of model compression, as far as we know. More specifically, in this paper, we focus on the study of theory of model compression for pretrained neural network models and then derive practical compression algorithms given the proposed theory.

% -----------------------------------------

\section{Rate distortion theory for model compression}
\label{sec:rate_distortion}

In this section, we briefly introduce the rate distortion theory for data compression. Then we extend the theory to compression of model parameters. 
%We use the minimum description length principle by~\cite{rissanen1978modeling} to bridge the gap between data compression and model compression. \todo{Still want to mention MDL?}

\subsection{Review of rate distortion theory for data compression}
Rate distortion theory, firstly introduced by~\citet{shannon1959coding} and further developed by \citet{berger1971rate}, is an important concept in information theory which gives theoretical description of lossy data compression. It addressed the minimum average number of $R$ bits, to transmit a random variable such that the receiver can reconstruct the random variable with distortion $D$. 

\begin{figure}
    \centering
    \includegraphics[width=0.45\textwidth]{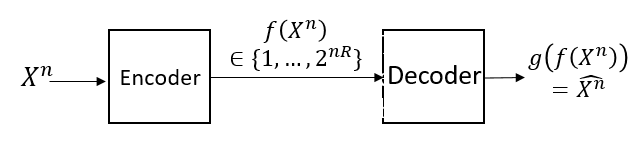}
    \caption{An illustration of encoder and decoder.}
    \label{fig:encoder_decoder}
\end{figure}

Precisely, let $X^n = \{X_1, X_2 \dots X_n\} \in \mathcal{X}^n$ be i.i.d. random variables from distribution $P_X$. An encoder $f_n: \mathcal{X}^n \to \{1, 2, \dots, 2^{nR}\}$ maps the message $X^n$ into codeword, and a decoder $g_n: \{1, 2, \dots, 2^{nR}\} \to \mathcal{X}^n$ reconstruct the message by an estimate $\hat{X}^n$ from the codeword. See Figure~\ref{fig:encoder_decoder} for an illustration.

A distortion function $d: \mathcal{X} \times \mathcal{X} \to \mathbb{R}^+$ quantifies the difference of the original and reconstructed message. Distortion between sequence $X^n$ and $\hat{X}^n$ is defined as the average distortion of $X_i$'s and $\hat{X}_i$'s. Commonly used distortion function includes Hamming distortion function $d(x, \hat{x}) = \mathbbm{1}[x \neq \hat{x}]$ for $\mathcal{X} = \{0, 1\}$ and square distortion function $d(x, \hat{x}) = (x - \hat{x})^2$ for $\mathcal{X} = \mathbb{R}$.

Now we are ready to define the rate-distortion function for data compression.

\begin{definition}
    A rate-distortion pair $(R,D)$ is {\em achievable} if there exists a series of (probabilistic) encoder-decoder $(f_n, g_n)$ such that the alphabet of codeword has size $2^{nR}$ and the expected distortion \\ $\lim_{n \to \infty} \mathbb{E} [d(X^n, g_n(f_n(X^n)))] \leq D$.
\end{definition}

\begin{definition}
    Rate-distortion function $R(D)$ equals to the infimum of rate $R$ such that rate-distortion pair $(R,D)$ is achievable.
\end{definition}

The main theorem of rate-distortion theory (\citet[Theorem 10.2.1]{cover2012elements}) states as follows,
\begin{theorem}
    Rate distortion theorem for data compression.
    \begin{eqnarray}
        R(D) &=& \min_{P_{\hat{X}|X}: \mathbb{E}[d(X, \hat{X})] \leq D} I(X; \hat{X}) \;.
    \end{eqnarray}
    \label{thm:rate_distortion}
\end{theorem}
The rate distortion quantifies the fundamental limit of data compression, i.e., {\em at least} how many bits are needed to compress the data, given the quality of the reconstructed data. Here is an example for rate-distortion function.

\begin{example}
    If $X \sim \mathcal{N}(0, \sigma^2)$, the rate distortion function is given by 
    \begin{eqnarray*}
    R(D) &=& \begin{cases} \frac{1}{2} \log_2 (\sigma^2 / D) \quad {\rm if\,} D \leq \sigma^2 \\ 0 \quad\quad\quad {\rm if\,} D > \sigma^2 \end{cases}.
    \end{eqnarray*}
\end{example}
If the required distortion $D$ is larger than the variance of the Gaussian variable $\sigma^2$, we simply transmit $\hat{X} = 0$; otherwise, we will transmit $\hat{X}$ such that $\hat{X} \sim \mathcal{N}(0, \sigma^2 - D)$, $X - \hat{X} \sim \mathcal{N}(0,D)$ where $\hat{X}$ and $X - \hat{X}$ are independent.

% \begin{figure}
%     \centering
%     \includegraphics[width=0.4\textwidth]{rate_distortion.PNG}
%     \put(-100,-10){$D$}
%     \put(-210,70){$R(D)$}
%     \caption{Rate distortion function for $X \sim \mathcal{N}(0,1)$.}
%     \label{fig:rate_distortion}
% \end{figure}

\subsection{Rate distortion theory for model compression}

Now we extend the rate distortion theory for data compression to model compression. To apply the rate distortion theory to model compression, we view the weights in the model as a multi-dimensional random variable $W \in \mathbb{R}^m$ following distribution $P_W$. The randomness comes from multiple sources including different distributions of training data, randomness of training data and randomness of training algorithm. The compressor can also be random hence we describe the compressor by a conditional probability $P_{\hat{W}|W}$. Now we define the distortion and rate in model compression, analogously to the data compression scenario.

{\bf Distortion.} Assume we have a neural network $f_w$ that maps input $x \in \mathbb{R}^{d_x}$ to $f_w(x)$ in output space $\mathcal{S}$. For regressors, $f_w(x)$ is defined as the output of the neural network on $\mathbb{R}^{d_y}$. Analogous to the square distortion in data compression, We define the distortion to be the expected $\ell_2$ distance between $f_w$ and $f_{\hat{w}}$, i.e.
    \begin{eqnarray}
        d(w, \hat{w}) \equiv \mathbb{E}_X \left[ \| f_w(X) - f_{\hat{w}}(X) \|_2^2 \right] \;.
        \label{eq:distortion_regression}
    \end{eqnarray}
For classfiers, $f_w(x)$ is defined as the output probability distribution over $C$ classes on the simplex $\Delta^{C-1}$. We define the distortion to be the expected distance between $f_w$ and $f_{\hat{w}}$, i.e.
    \begin{eqnarray}
        d(w, \hat{w}) \equiv \mathbb{E}_X \left[\, D(f_{\hat{w}}(X) || f_w(X)) \,\right] \;.
        \label{eq:distortion_classification}
    \end{eqnarray}

Here $D$ could be any statistical distance, including KL divergence, Hellinger distance, total variation distance, etc. Such a definition of distortion captures the difference between the original model and the compressed model, averaged over data $X$, and measures the quality of a compression algorithm.
  
{\bf Rate.} In data compression, the rate is defined as the description length of the bits necessary to communicate the compressed data $\hat{X}$. The compressor outputs $\hat{X}$ from a finite {\em code book} $\mathcal{X}$. The description consists the {\em code word} which are the indices of $\hat{x}$ in the code book, and the description of the {\em code book}. 
    
In rate distortion theory, we ignore the code book length. Since we are transmitting a sequence of data $X^n$, the code word has to be transmitted for each $X_i$ but the code book is only transmitted once. In asymptotic setting, the description length of code book can be ignored, and the rate is defined as the description length of the  code word.
    
In model compression, we also define the rate as the code word length, by assuming that an underlying distribution $P_W$ of the parameters exists and infinitely many models whose parameters are i.i.d. from $P_W$ will be compressed. In practice, we only compress the parameters once so there is no distribution of the parameters. Nevertheless, the rate distortion theory can also provide important intuitions for one-time compression, explained in Section~\ref{sec:relu}.

Now we can define the rate distortion function for model compression. Analogously to Theorem~\ref{thm:rate_distortion}, the rate distortion function for model compression is defined as follows,
\begin{definition}
    Rate distortion function for model compression.
    \begin{eqnarray}
    R(D) = \min_{P_{\hat{W}|W}: \mathbb{E}_{W, \hat{W}} \left[ d(W,\hat{W}) \right] \leq D } I(W; \hat{W}).
\end{eqnarray}
\label{def:rate_distortion}
\end{definition}

In the following sections we establish a lower bound of the rate-distortion function.

% ----------------------------------------- 
\section{Lower bound and achievability for rate distortion function}
\label{sec:lower_bound}
In this section, we study the lower bound for rate distortion function in Definition~\ref{def:rate_distortion}. We provide a lower bound for the rate distortion function, and prove that this lower bound is achivable for linear regression models.

\subsection{Lower bound for linear model}\label{sec:achieve}

Assume that we are going to compress a linear regression model $f_w(x) = w^T x$. We assume that the mean of data $x \in \mathbb{R}^m$ is zero and the covariance matrix is diagonal, i.e., $\mathbb{E}_X [X_i^2] = \lambda_{x,i} > 0$ and $\mathbb{E}_X [X_i X_j] = 0$ for $i \neq j$. Furthermore, assume that the parameters $W \in \mathbb{R}^m$ are drawn from a Gaussian distribution $\mathcal{N}(0, \Sigma_W)$. The following theorem gives the lower bound of the rate distortion function for the linear regression model.

\begin{theorem}
The rate-distortion function of the linear regression model $f_w(x) = w^T x$ is lower bounded by
\begin{eqnarray*}
		R(D) \geq \underline{R}(D) = \frac{1}{2} \log \det (\Sigma_W) - \sum_{i=1}^m \frac{1}{2} \log (D_i),
\end{eqnarray*}
where
\begin{eqnarray*}
	D_i = \begin{cases}
    	\mu/\lambda_{x,i} \quad {\rm if} \mu < \lambda_{x,i} \mathbb{E}_W [W_i^2] \;, \\
        \mathbb{E}_W [W_i^2]  \quad {\rm if} \mu \geq \lambda_{x,i}\mathbb{E}_W [W_i^2]  \;,
    \end{cases}
\end{eqnarray*}
where $\mu$ is chosen that $\sum_{i=1}^m \lambda_{x,i} D_i = D$.
\label{thm:linear_lower_bound}
\end{theorem}

This lower bound gives rise to a ``weighted water-filling'' approach, which differs from the classical ``water-filling'' for rate distortion of colored Gaussian source in~\citet[Figure 13.7]{cover2012elements}. The details and graphical explanation of the ``weighted water-filling'' can be found in Appendix~\ref{sec:append_lower_bound}.

\subsection{Achievability}

We show that, the lower bound give in Theorem~\ref{thm:linear_lower_bound} is achievable. Precisely, we have the following theorem.

\begin{theorem}
    There exists a class of probabilistic compressors $P_{\hat{W}^*|W}^{(D)}$ such that $\mathbb{E}_{P_W \circ P_{\hat{W}^*|W}^{(D)}} \left[ d(W, \hat{W}^*) \right] = D$ and $I(W; \hat{W}^*) = \underline{R}(D)$.
\end{theorem}
The optimal compressor is Algorithm~\ref{algo:optimal} in Appendix~\ref{sec:append_lower_bound}. Intuitively, the optimal compressor does the following
\begin{itemize}
    \item Find the optimal water levels $D_i$ for ``weighted water filling'', such that the expected distortion $D = \mathbb{E}_{W, \hat{W}} [ d(W, \hat{W}) ] = \mathbb{E}_{W, \hat{W}} [\hat{W}^T \Sigma_X (W - \hat{W})]$ is minimized given certain rate. 
    \item Add a noise $Z_i$ which is independent of $\hat{W}_i=W_i+Z_i$ and has a variance proportional to the water level. That is possible since $W$ is Gaussian.
\end{itemize}
We can check that the compressor makes all the inequalities become equality, hence achieve the lower bound. The full proof of the lower bound and achievability can be found in Appendix~\ref{sec:append_lower_bound}.

\section{Improved objective for model compression}
\label{sec:relu}
In the previous sections, we study the rate-distortion theory for model compression. In rate-distortion theory, we assume that there exists a prior distribution $P_W$ on the weights $W$, and prove the tightness of the lower bound in the asymptotic scenario. However, in practice, we only compress one particular pre-trained model, so there are no prior distribution of $W$. Nonetheless, we can still learn something important from the achivability of the lower bound, by extracting two ``golden rules'' from the optimal algorithm for linear regression. 

\subsection{Two golden rules}
Recall that for linear regression model, to achieve the smallest rate given certain distortion (or, equivalently, achieve the smallest distortion given certain rate), the optimal compressor need to do the following: (1) find appropriate ``water levels'' such that the expected distortion $E_{W, \hat{W}} [d(W, \hat{W})] = \mathbb{E}_{W, \hat{W}, X} [(W^T X - \hat{W}^T X )^2] = \mathbb{E}_{W, \hat{W}} [(W - \hat{W})^T \Sigma_X (W - \hat{W})]$ is minimized. (2) make sure that $\hat{W}_i$ is independent with $W_i - \hat{W}_i$, in other words, $\mathbb{E}_{W, \hat{W}}[\hat{W}^T \Sigma_X (W - \hat{W})] = 0$. Hence, we extract the following two ``golden rules'':

\begin{enumerate}
    \item $\mathbb{E}_{W, \hat{W}} [\hat{W}^T \Sigma_X (W - \hat{W})] = 0$
    \item $\mathbb{E}_{W, \hat{W}} [(W - \hat{W})^T \Sigma_X (W - \hat{W})]$ should be minimized, given certain rate.
\end{enumerate}

For practical model compression, we adopt these two ``golden rules'', by making the following amendments. First, we discard the expectation over $W$ and $\hat{W}$ since there is only one model to be compressed. Second, the distortion can be written as $d(w, \hat{w}) = (w - \hat{w})^T \Sigma_X (w - \hat{w})$ only for linear models. For non-linear models, the distortion function is complicated, but can be approximated by a simpler formula. For non-linear regression models, we take first order Taylor expansion of the function $f_{\hat{w}}(x) \approx f_w(x) + (\hat{w}-w)^T \nabla_w f_w(x)$, and have
\begin{eqnarray}
    d(w, \hat{w}) &=& \mathbb{E}_X \left[\, \|f_w(X) - f_{\hat{w}}(X)\|_2^2\,\right] \,\notag\\
    &\approx& \mathbb{E}_X \left[ (w - \hat{w})^T \nabla_w f_w(X) (\nabla_w f_w(X))^T (w - \hat{w}) \right] \,\notag\\
    &=& (w - \hat{w})^T I_w (w - \hat{w}) \notag
\end{eqnarray}
where the ``weight importance matrix'' defined as
\begin{eqnarray}
    I_w = \mathbb{E}_X \left[ \nabla_w f_w(X) (\nabla_w f_w(X))^T\right],
    \label{eq:wim_regression}
\end{eqnarray} 
quantifies the relative importance of each weight to the output. For linear regression models, weight importance matrix $I_w$ equals to $\Sigma_X$.

For classification models, we will first approximate the KL divergence. Using the Taylor expansion $x \log (x/a) \approx (x-a) + (x-a)^2/(2a)$ for $x/a \approx 1$, the KL divergence $D_{KL}(P||Q)$ for can be approximated by $ D_{KL}(P||Q) \approx \sum_{i} (P_i - Q_i) + (P_i - Q_i)^2/(2P_i) = \sum_i  (P_i - Q_i)^2/(2P_i)$, or in vector form $D_{KL}(P||Q) \approx \frac{1}{2}(P-Q)^T {\rm diag}[P^{-1}](P-Q)$. Therefore,
\begin{eqnarray}
    d(w, \hat{w}) &=& \mathbb{E}_X \left[ D_{KL} (f_{\hat{w}}(X) || f_{w}(X)) \right] \,\notag\\
    &\approx& \frac{1}{2} \mathbb{E}_X \big[ (f_w(X) - f_{\hat{w}}(X))^T  {\rm diag}[f_w^{-1}(X)]  (f_w(X) - f_{\hat{w}}(X))  \big] \,\notag\\
    &\approx& \frac{1}{2} \mathbb{E}_X \big[ (w - \hat{w})^T (\nabla_w f_w(X)) {\rm diag}[f_w^{-1}(X)]  (\nabla_w f_w(X))^T (w - \hat{w}) \big]. \notag
\end{eqnarray}
So the weight importance matrix is given by
\begin{eqnarray}
    I_w = \mathbb{E}_X \left[ (\nabla_w f_w(X)) {\rm diag}[f_w^{-1}(X)] (\nabla_w f_w(X))^T \right].
    \label{eq:wim_classification}
\end{eqnarray}
This weight importance matrix is also valid for many other statistical distances, including reverse KL divergence, Hellinger distance and Jenson-Shannon distance.

Now we define the two ``golden rules'' for practical model compression algorithms,
\begin{enumerate}
    \item $\hat{w}^T I_w (w - \hat{w}) = 0$,
    \item $(w - \hat{w})^T I_w (w - \hat{w})$ is minimized given certain constraints.
\end{enumerate}
In the following subsection we will show the optimality of the ``golden rules'' for a one-hidden-layer neural network.

\subsection{Optimality for one-hidden-layer ReLU network}
We show that if a compressor of a one-hidden-layer ReLU network satisfies the two ``golden rules'', it will be the optimal compressor, with respect to mean-square-error. Precisely, consider the one-hidden layer ReLU neural network $f_w(x) = ReLU(w^T x)$, where the distribution of input $x \in \mathbb{R}^m$ is $\mathcal{N}(0, \Sigma_X)$. Furthermore, we assume that the covariance matrix $\Sigma_X = {\rm diag}[\lambda_{x,1}, \dots, \lambda_{x,m}]$ is diagonal and $\lambda_{x,i} > 0$ for all $i$. We have the following theorem.
\begin{theorem}
    If compressed weight $\hat{w}^*$ satisfies $\hat{w}^* I_w (\hat{w}^*-w) = 0$ and 
    \begin{eqnarray*}
        \hat{w}^* = \arg\min_{\hat{w} \in \hat{\mathcal{W}}} (w - \hat{w})^T I_w (w - \hat{w}),
    \end{eqnarray*}
    where $\hat{\mathcal{W}}$ is some class of compressors, then
    \begin{eqnarray*}
        \hat{w}^* = \arg\min_{\hat{w} \in \hat{\mathcal{W}}} \mathbb{E}_X \left[ (f_w(X) - f_{\hat{w}}(X))^2 \right].
    \end{eqnarray*}
    \label{thm:optimal_relu}
\end{theorem}
The proof uses the techniques of Hermite polynomials and Fourier analysis on Gaussian spaces, inspired by~\citet{ge2017learning}. The full proof can be found in Appendix~\ref{sec:append_proof}.

Here $\hat{\mathcal{W}}$ denotes a class of compressors, with some constraints. For example, $\hat{\mathcal{W}}$ could be the class of pruning algorithms where no more than 50\% weights are pruned, or $\hat{\mathcal{W}}$ could be the class of quantization algorithm where each weight is quantized to 4 bits. Theoretically, it is not guaranteed that the two ``golden rules'' can be satisfied simultaneously for every $\hat{\mathcal{W}}$, but in the following subsection we show that they can be satisfied simultaneously for two of the most commonly used class of compressors --- pruning and quantization. Hence, minimizing the objective $(w - \hat{w})^T I_w (w - \hat{w})$ will be optimal for pruning and quantization.
 
\subsection{Improved objective for pruning and quantization}
Pruning and quantization are two most basic and useful building blocks of modern model compression algorithms, For example, DeepCompress~\cite{han2015deep} iteratively prune, retrain and quantize the neural network and achieve state-of-the-art performances on large neural networks. 

In pruning algorithms, we choose a subset $S \in [m]$ and set $\hat{w}_i = 0$ for all $i \in S$ and $\hat{w}_i = w_i$ for $i \not\in S$. The compression ratio is evaluated by the proportion of unpruned weights $r = (m - |S|)/m$. Since either $\hat{w}_i$ or $w_i - \hat{w}_i$ is zero, so the first ``golden rule'' is automatically satisfied, so we have the following corollary.

\begin{corollary}
For any fixed $r$, let 
\begin{align*}
\hat{w}^*_r = \arg\min_{S: \frac{d-|S|}{d} = r} (w - \hat{w})^T I_w (w - \hat{w}), 
\end{align*}
Then 
\begin{align*}
    \hat{w}^*_r = \arg\min_{S: \frac{d-|S|}{d} = r} \mathbb{E}_X \left[ (f_w(X) - f_{\hat{w}}(X))^2 \right]. 
\end{align*}
\end{corollary}
In quantization algorithms, we cluster the weights into $k$ centroids $\{c_1, \dots, c_k\}$. The algorithm optimize the centroids as long as the assignments of each weight $A_i \in [k]$. The final compressed weight is given by $\hat{w}_i = c_{A_i}$. Usually $k$-means algorithm are utilized to minimize the centroids and assignments alternatively. The compression ratio of quantization algorithm is given by 
\begin{eqnarray}
r = \frac{mb}{m \sum_{j=1}^k \frac{m_j}{m} \lceil \log_2 \frac{m}{m_j} \rceil + kb}, \notag
\end{eqnarray}
where $m$ is the number of weights and $b$ is the number of bits to represent one weight before quantization (usually 32). By using Huffman coding, the average number of bits for each weight is given by $\sum_{j=1}^k (m_j/m) \lceil \log_2 (m/m_j) \rceil$, where $m_j$ is the number of weights assigned to the $j$-th cluster.

If we can find the optimal quantization algorithm with respect to $(w-\hat{w})^T I_w (w-\hat{w})$, then each centroids $c_j$ should be optimal, i.e.
\begin{eqnarray*}
    0 = \frac{\partial}{\partial c_j} (w - \hat{w})^T I_w (w - \hat{w}) = - 2 \left(\sum_{i: A_i=j} e_i^T\right) I_w (w - \hat{w})
\end{eqnarray*}
where $e_i$ is the $i$-th standard basis. Therefore, we have
\begin{eqnarray*}
    \hat{w} I_w (\hat{w}-w) = \left( \sum_{j=1}^k c_j (\sum_{i: A_i=j} e_i) \right)^T I_w (w - \hat{w}) = \sum_{j=1}^k c_j \left( (\sum_{i: A_i=j} e_i^T) I_w (w - \hat{w}) \right) = 0.
\end{eqnarray*}
Hence the first ``golden rule'' is satisfied if the second ``golden rule'' is satisfied. So we have

\begin{corollary}
For any fixed number of centroids $k$, let 
\begin{eqnarray*}
\hat{w}^*_k = \arg\min_{\{c_1, \dots, c_k\}, A \in [k]^m} (w - \hat{w})^T I_w (w - \hat{w}),
\end{eqnarray*}
then 
\begin{eqnarray*}
    \hat{w}^*_k = \arg\min_{\{c_1, \dots, c_k\}, A \in [k]^m} \mathbb{E}_X \left[ (f_w(X) - f_{\hat{w}}(X))^2 \right].
\end{eqnarray*}
\end{corollary}

As corollaries of Theorem~\ref{thm:optimal_relu}, we proposed to use $(w - \hat{w})^T I_w (w - \hat{w})$ as the objective for pruning and quantization algorithms, which can achieve the minimum MSE for one-hidden-layer ReLU neural network.

\section{Experiments}
\label{sec:experiment}
In the previous section, we proved that a pruning or quantization algorithm that minimizes the objective $(w - \hat{w})^T I_w (w - \hat{w})$ also minimizes the MSE loss for one-hidden-layer ReLU neural network. In this section, we show that this objective can also improve pruning and quantization algorithm for larger neural networks on real data.\footnote{We leave combinations of pruning, model retraining and quantization like~\citet{han2015deep} as future work.}

We test the objectives on the following neural network and datasets. 
\begin{enumerate*}
    \item 3-layer fully connected neural network on MNIST.
    \item Convolutional neural network with 5 convolutional layers and 3 fully connected layers on CIFAR 10 and CIFAR 100.
\end{enumerate*}
We load the pretrained models from~\url{https://github.com/aaron-xichen/pytorch-playground}.

In Section~\ref{sec:experiment}.1, we use the weight importance matrix for classification in Eq.~\eqref{eq:wim_classification}, which is derived by approximating the distortion of KL-divergence. This weight importance matrix does not depend on the training labels, so the induced pruning/quantization algorithms is called ``unsupervised compression''. Furthermore, if the training labels are available, we treat the loss function $\mathcal{L}_w(X,Y): \mathcal{X} \times \mathcal{Y} \to \mathbb{R}^+$ as the function to be compressed, and derive several pruning/quantization objectives. The induced pruning/quantization methods are called ``supervised compression'' and are studied in Section~\ref{sec:experiment}.2.

%\todo{It would be still confusing using 'unsupervised experiments', since the model was trained with labels. any thoughts? How about 'unsupervised compression experiments' and 'supervised compression experiments'.}

\subsection{Unsupervised Compression Experiments}
Recall that for classification problems, the weight importance matrix is defined as
\begin{eqnarray*}
I_w &=& \mathbb{E}_X \left[ \nabla_w f_w(X) {\rm diag}[f_w^{-1}(X)](\nabla_w f_w(X))^T\right].
\end{eqnarray*}
For computational simplicity, we drop the off-diagonal terms of $I_w$, and simplify the objective to \\ $\sum_{i=1}^m \mathbb{E}_X [\frac{(\nabla_{w_i} f_w(X))^2}{f_w(X)}] (w_i - \hat{w}_i)^2$. To minimize the proposed objective, a pruning algorithm just prune the weights with smaller $\mathbb{E}_X [\frac{(\nabla_{w_i} f_w(X))^2}{f_w(X)}] w_i^2$ greedily. A quantization algorithm uses the weighted $k$-means algorithm~\cite{choi2016towards} to find the optimal centroids and assignments. We compare the proposed objective with the baseline objective $\sum_{i=1}^m (w_i - \hat{w}_i)^2$, which were used as building blocks in DeepCompress~\cite{han2015deep}. We compare the objectives in Table~\ref{table:objective_1}.

\begin{table}[htbp]
\centering
\begin{tabular}{c|c}
\hline
Name &  Minimizing objective \\ \hline
Baseline & $\sum_{i=1}^m  (w_i - \hat{w}_i)^2$ \\ \hline
Proposed & $\sum_{i=1}^m \mathbb{E}_X [\frac{(\nabla_{w_i} f_w(X))^2}{f_w(X)}] (w_i - \hat{w}_i)^2$ \\ \hline
\end{tabular}
\label{table:objective_1}
\caption{Comparison of unsupervised compression objectives.}
\end{table}

For pruning experiment, we choose the same compression rate for every convolutional layer and fully-connected layer, and plot the test accuracy and test cross-entropy loss against compression rate. For quantization experiment, we choose the same number of clusters for every convolutional and fully-connected layer.  Also we plot the test accuracy and test cross-entropy loss against compression rate.
To reduce the variance of estimating the weight importance matrix $I_w$, we use the {\em temperature scaling} method introduced by~\citet{guo2017calibration} to improve model calibration. 

\begin{figure}[htb]
    \centering
	\includegraphics[width=0.4\textwidth]{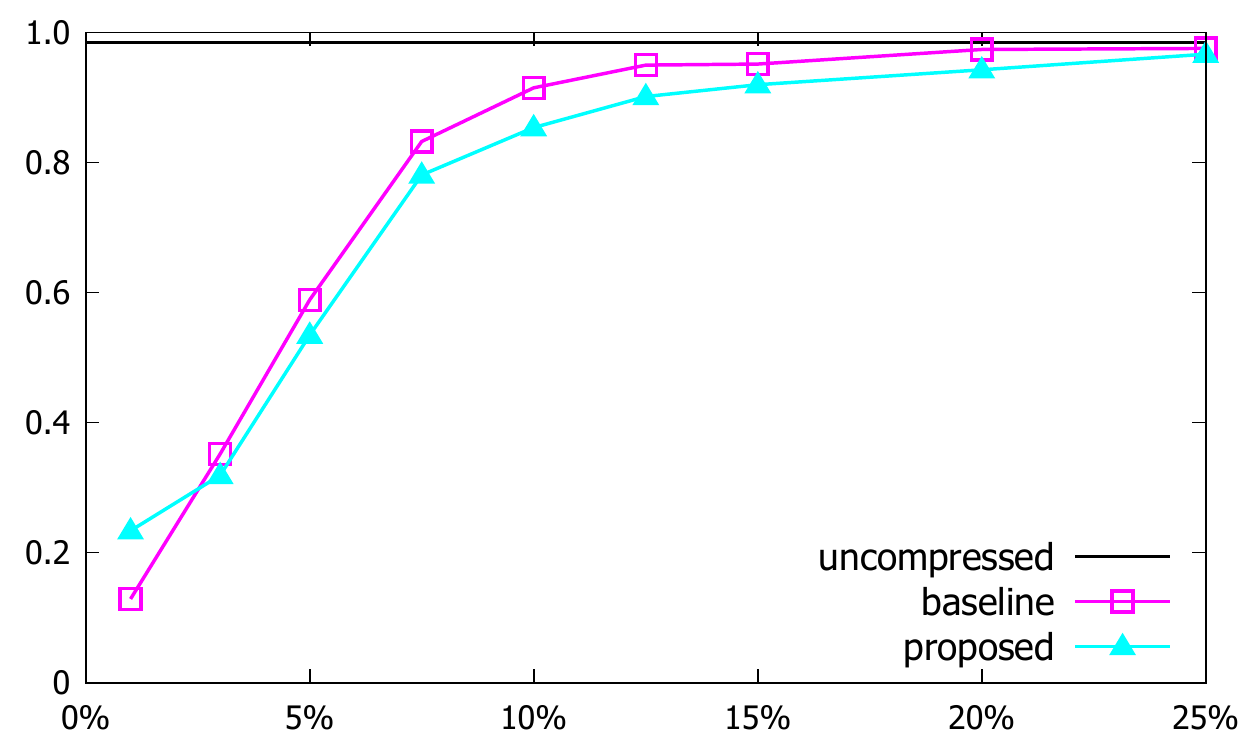}
	\put(-200,40){\rotatebox{90}{Accuracy}}
    \includegraphics[width=0.4\textwidth]{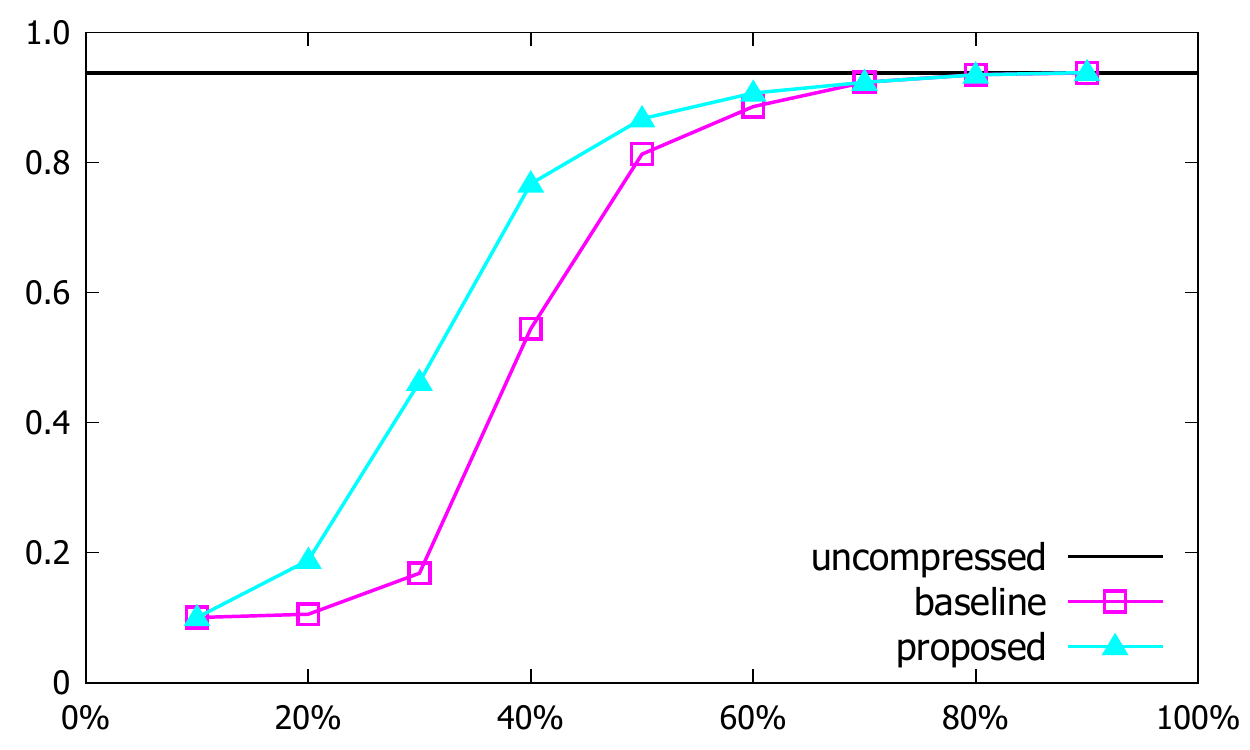}
    \includegraphics[width=0.4\textwidth]{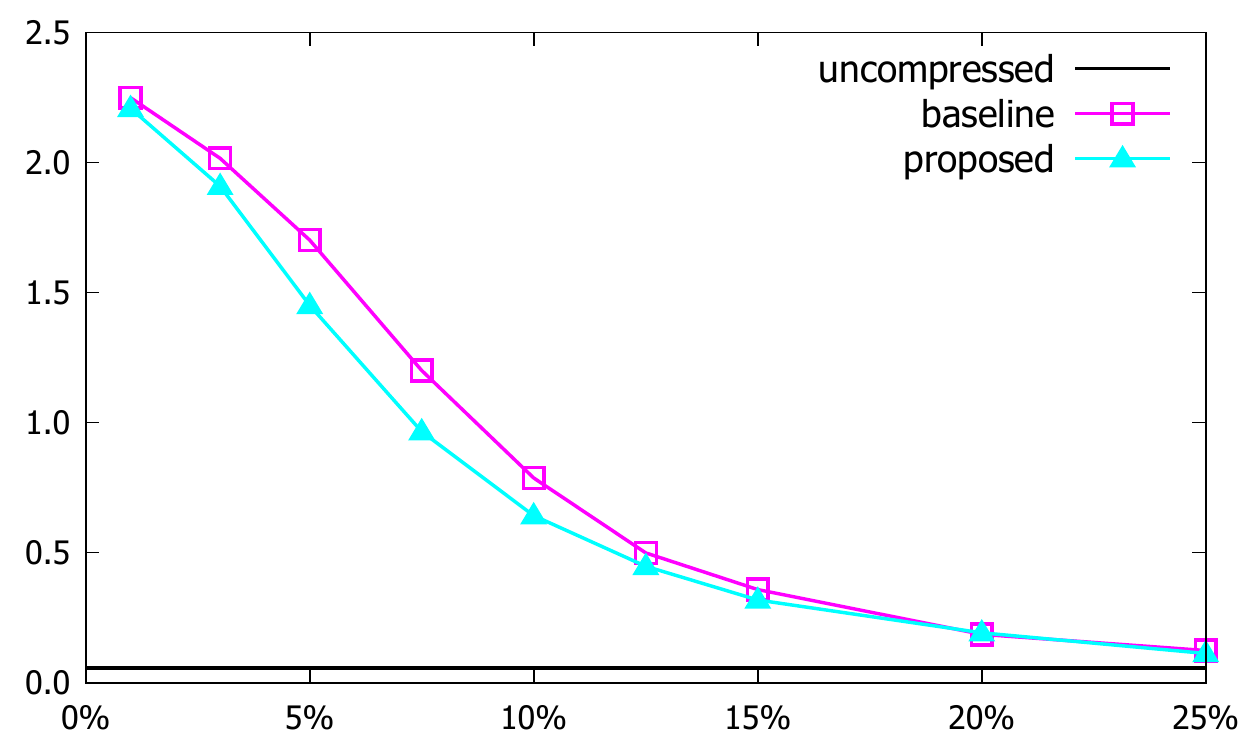}
    \put(-200,30){\rotatebox{90}{Cross Entropy}}
    \includegraphics[width=0.4\textwidth]{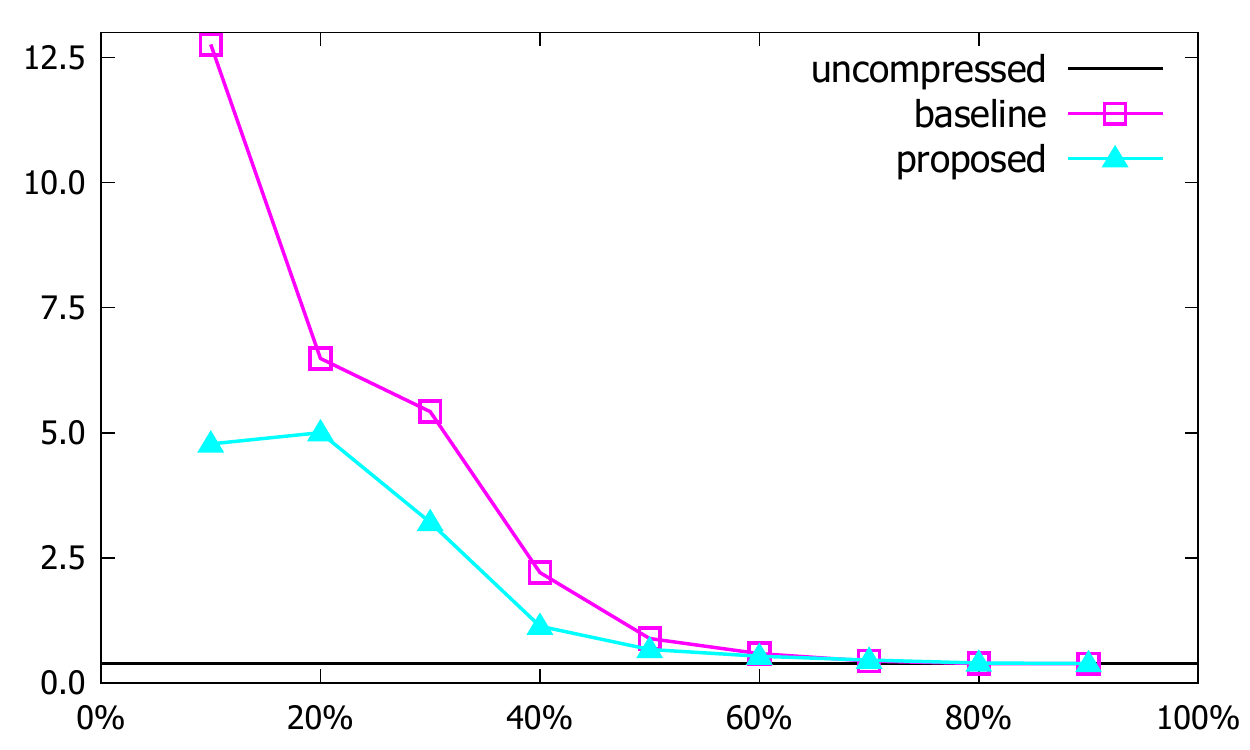}
    \put(-130,-10){Compression Ratio}
    \put(-320,-10){Compression Ratio}
    \caption{Result for unsupervised pruning experiment. Left: fully-connected neural network on MNIST (Top: test accuracy, Bottom: test cross entropy loss). Right: convolutional neural network on CIFAR10 (Top: test accuracy, Bottom: test cross entropy loss).}
    \label{fig:KL_pruning}
\end{figure}

\begin{figure}[htb]
    \centering
	\includegraphics[width=0.4\textwidth]{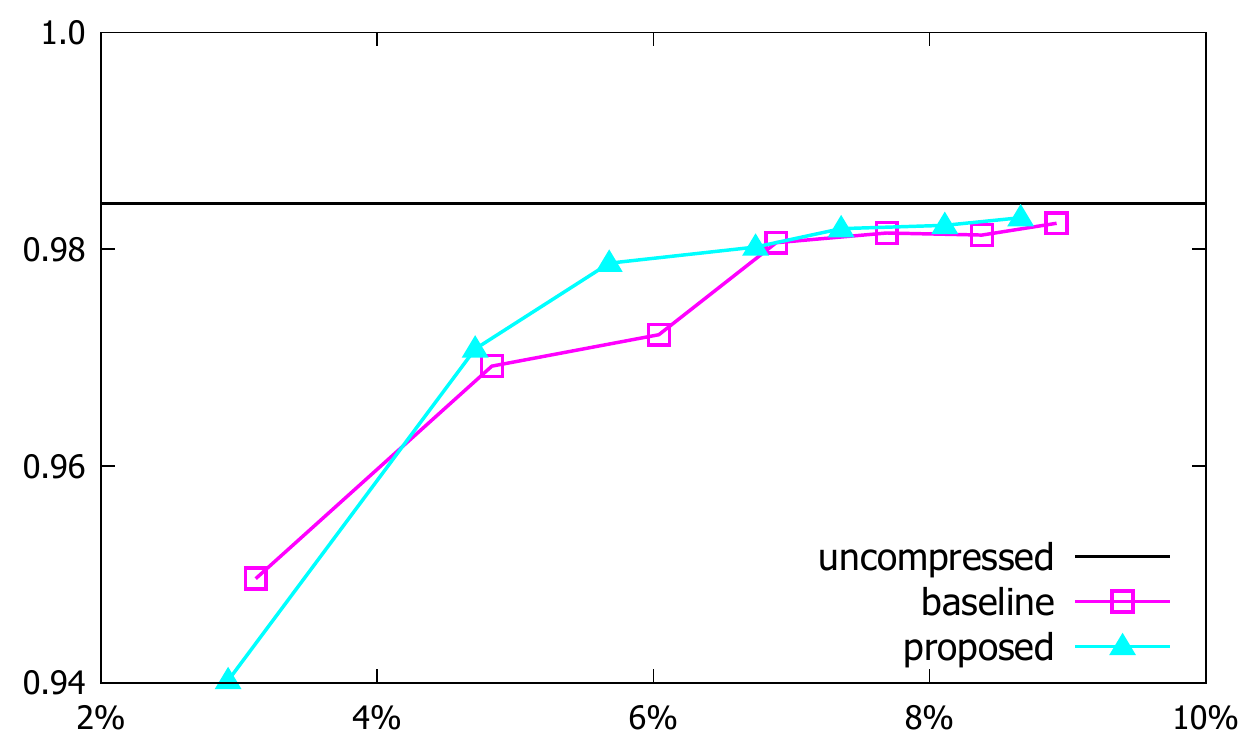}
	\put(-200,40){\rotatebox{90}{Accuracy}}
    \includegraphics[width=0.4\textwidth]{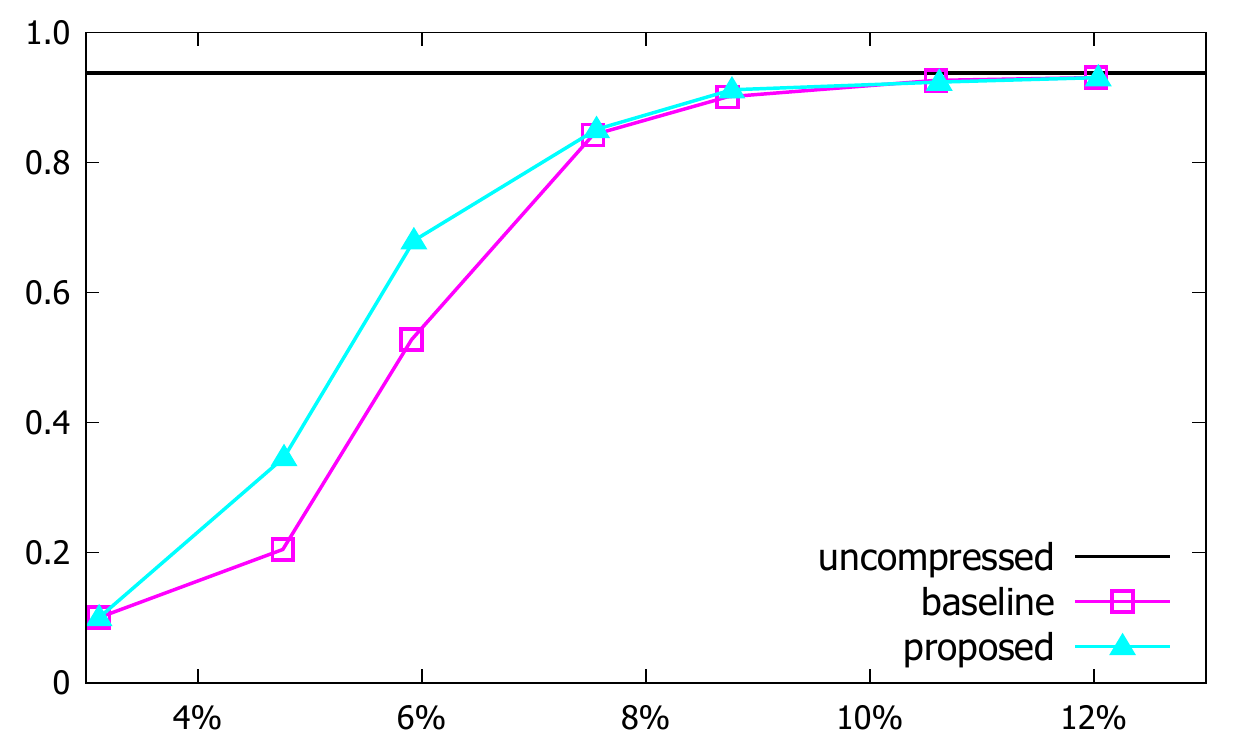}
    \includegraphics[width=0.4\textwidth]{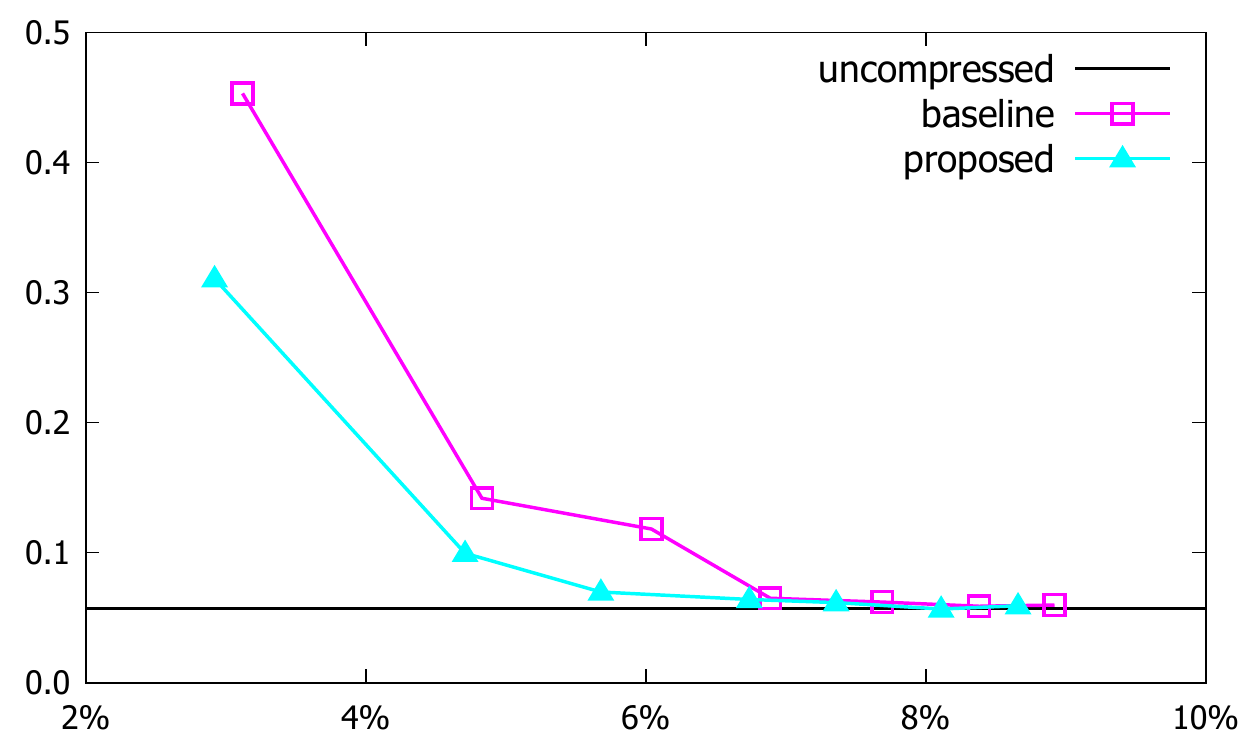}
    \put(-200,30){\rotatebox{90}{Cross Entropy}}
    \includegraphics[width=0.4\textwidth]{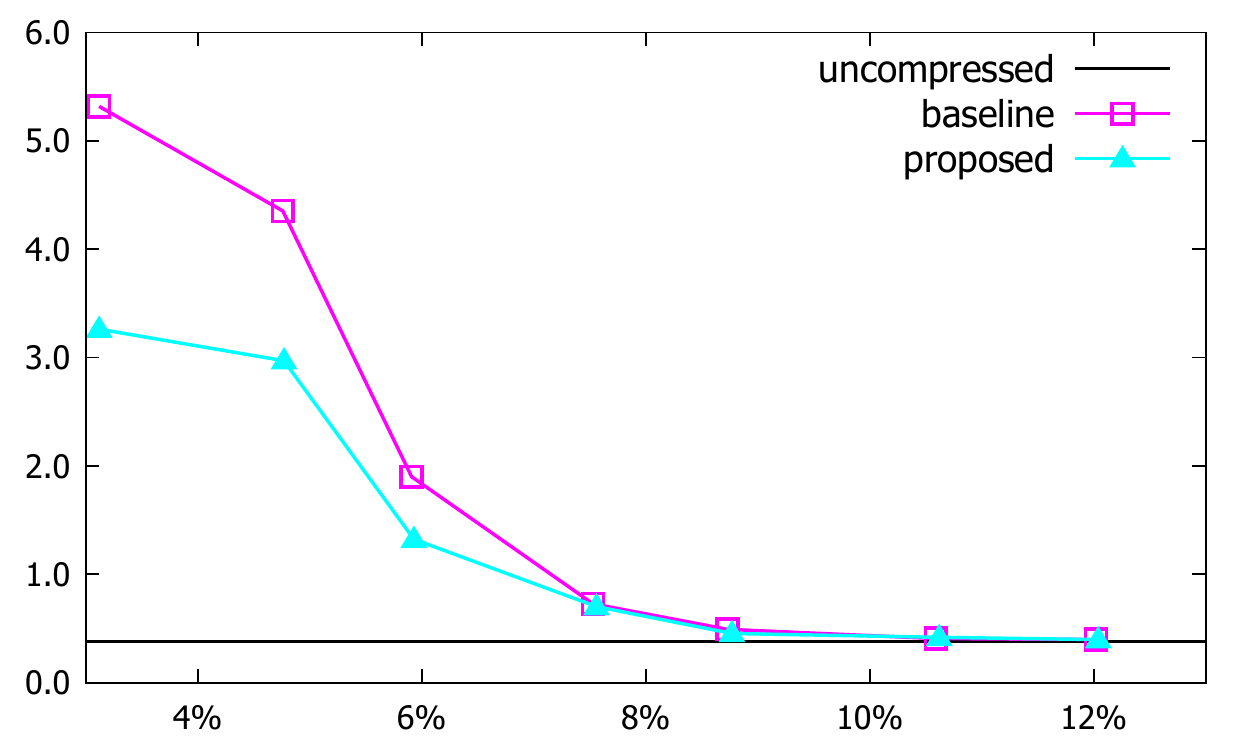}
    \put(-130,-10){Compression Ratio}
    \put(-320,-10){Compression Ratio}
    \caption{Result for unsupervised quantization experiment. Left: fully-connected neural network on MNIST (Top: test accuracy, Bottom: test cross entropy loss). Right: convolutional neural network on CIFAR10 (Top: test accuracy, Bottom: test cross entropy loss).}
    \label{fig:KL_quantization}
\end{figure}

We show that results of pruning experiment in Figure~\ref{fig:KL_pruning}, and the results of quantization experiment in Figure~\ref{fig:KL_quantization}. We can see that the proposed objective gives better validation cross-entropy loss than the baseline, for every different compression ratios. The proposed objective also gives better validation accuracy in most scenarios. We relegate the results for CIFAR100 in Appendix~\ref{sec:append_experiment}.

\subsection{Supervised Compression Experiments}

In the previous experiment, we only use the training data to compute the weight importance matrix. But if we can use the training label as well, we can further improve the performance of pruning and quantization algorithms. If the training label is available, we can view the cross-entropy loss function $\mathcal{L}(f_w(x), y) = \mathcal{L}_w(x, y)$ as a function from $\mathcal{X} \times \mathcal{Y} \to \mathbb{R}^+$, and define the distortion function as 
\begin{eqnarray*}
    d(w, \hat{w}) &=& \mathbb{E}_{X,Y} \left[ (\mathcal{L}_w(X,Y) - \mathcal{L}_{\hat{w}}(X,Y))^2 \right].
\end{eqnarray*}
Taking first order approximation of the loss function gives the supervised weight importance matrix,
\begin{eqnarray*}
    I_w &=& \mathbb{E} \left[ \nabla_w \mathcal{L}_w(X,Y) (\nabla_w \mathcal{L}_w(X,Y))^T\right].
\end{eqnarray*}

We write $\mathbb{E}$ instead of $\mathbb{E}_{X,Y}$ for simplicity. Similarly, we drop the off-diagonal terms for ease of computation, and simplify the objective to $\sum_{i=1}^m \mathbb{E} [(\nabla_{w_i} \mathcal{L}_w(X,Y))^2] (w_i - \hat{w}_i)^2$, which is called gradient-based objective. Note that for well-trained model, the expected value of gradient $\mathbb{E} [\nabla_w \mathcal{L}_w(X,Y)]$ is closed to zero, but the second moment of the gradient $\mathbb{E} [\nabla_w \mathcal{L}_w(X,Y) (\nabla_w \mathcal{L}_w(X,Y))^T]$ could be large. We compare this objective with the baseline objective $\sum_{i=1}^m (w_i - \hat{w}_i)^2$. We also compare with the hessian-based objective $\sum_{i=1}^m \mathbb{E} [\nabla^2_{w_i} \mathcal{L}_w(X,Y)] (w_i  -\hat{w}_i)^2$, which is used in~\cite{lecun1990optimal} and~\cite{hassibi1993second} for network pruning and~\cite{choi2016towards} for network quantization.  To estimate the diagonal entries of the Hessian matrix of the loss function with respect to the model parameters, we implemented Curvature Propagation~\cite{martens2012estimating} treating each layer and activation as a node.  The running time is proportional to the running time of the usual gradient back-propagation by a factor that does not depend on the size of the model.  Manually optimizing the local Hessian calculation at each node reduces memory usage and allows us to use larger batch size and larger number of samples for more accurate estimates. 

Furthermore, if we take second order approximation of the loss function, and drop the off-diagonal terms of the squared gradient matrix and squared hessian tensor, we have the following approximation
%\begin{eqnarray}
%     &&d(w, \hat{w}) = \mathbb{E} \big[ (\mathcal{L}_w(X,Y) - \mathcal{L}_{\hat{w}}(X,Y))^2 \big] \,\notag\\
%     &\approx& \mathbb{E} \big[ (\nabla_w \mathcal{L}_w(X,Y)^T (w - \hat{w}) \,\notag\\
%     &+& \frac{1}{2}(w - \hat{w})^T \nabla^2_w \mathcal{L}_w(X,Y) (w - \hat{w}))^2 \big] \,\notag\\
%     &=& \langle \mathbb{E} \left[ \nabla_w \mathcal{L}_w(X,Y) \otimes \nabla_w \mathcal{L}_w(X,Y) \right], (w -  \hat{w})^{\otimes 2} \rangle \,\notag\\
%     &+& \langle \mathbb{E} \left[ \nabla_w \mathcal{L}_w(X,Y) \otimes \nabla^2_w \mathcal{L}_w(X,Y) \right], (w -  \hat{w})^{\otimes 3} \rangle \,\notag\\
%     &+& \frac{1}{4} \langle \mathbb{E} \left[ \nabla^2_w \mathcal{L}_w(X,Y) \otimes \nabla^2_w \mathcal{L}_w(X,Y) \right], (w -  \hat{w})^{\otimes 4} \rangle \,\notag
% \end{eqnarray}
% here $\otimes$ denotes tensor outer production, and $\langle A, B \rangle = \sum_{i_1, ... i_j} A_{i_1, ...i_j} B_{i_1, ...i_j}$ denotes inner production of two tensors. For computational simplicity, we drop the off-diagonal terms of $\mathbb{E} \left[ \nabla_w \mathcal{L}_w(X,Y) \otimes \nabla_w \mathcal{L}_w(X,Y) \right]$ and $\mathbb{E} \left[ \nabla^2_w \mathcal{L}_w(X,Y) \otimes \nabla^2_w \mathcal{L}_w(X,Y) \right]$, and ignore the third order tensor $\mathbb{E} \left[ \nabla_w \mathcal{L}_w(X,Y) \otimes \nabla^2_w \mathcal{L}_w(X,Y) \right]$. So we have 
\begin{eqnarray}
    d(w, \hat{w}) &=& \mathbb{E} \big[ (\mathcal{L}_w(X,Y) - \mathcal{L}_{\hat{w}}(X,Y))^2 \big] \,\notag\\
    &\approx& \mathbb{E} \big[ (\nabla_w \mathcal{L}_w(X,Y)^T (w - \hat{w}) + \frac{1}{2}(w - \hat{w})^T \nabla^2_w \mathcal{L}_w(X,Y) (w - \hat{w}))^2 \big] \,\notag\\
    &\approx& \sum_{i=1}^m \mathbb{E} [(\nabla_{w_i} \mathcal{L}_w(X,Y))^2] (w_i - \hat{w}_i)^2 + \frac{1}{4} \sum_{i=1}^m \mathbb{E} [(\nabla^2_{w_i} \mathcal{L}_w(X,Y))^2] (w_i - \hat{w}_i)^4, \notag
\end{eqnarray}
which is called gradient+hessian based objective. For pruning algorithm, we can prune the weights with smaller $ \mathbb{E} [(\nabla_{w_i} \mathcal{L}_w(X,Y))^2] w_i^2 + \frac{1}{4} \mathbb{E} [(\nabla^2_{w_i} \mathcal{L}_w(X,Y))^2] w_i^4$ greedily. For quantization algorithm, we use an alternatice minimization algorithm in Appendix~\ref{sec:append_experiment} to find the minimum. We conclude the different supervised objectives in Table~\ref{table:objective_2}.

\begin{table}[htbp]
\centering
\begin{tabular}{c|c}
\hline
Name &  Minimizing objective \\ \hline
Baseline & $\sum_{i=1}^m  (w_i - \hat{w}_i)^2$ \\ \hline
Gradient & $\sum_{i=1}^m \mathbb{E} [(\nabla_{w_i} \mathcal{L}_w(X,Y))^2] (w_i - \hat{w}_i)^2$ \\ \hline
Hessian & $\sum_{i=1}^m \mathbb{E} [\nabla^2_{w_i} \mathcal{L}_w(X,Y)] (w_i  -\hat{w}_i)^2$ \\ \hline
Gradient & $\sum_{i=1}^m \mathbb{E} [(\nabla_{w_i} \mathcal{L}_w(X,Y))^2] (w_i - \hat{w}_i)^2$ \\
+ Hessian & $+ \frac{1}{4} \sum_{i=1}^m \mathbb{E} [(\nabla^2_{w_i} \mathcal{L}_w(X,Y))^2] (w_i - \hat{w}_i)^4 $ \\ \hline
\end{tabular}
\label{table:objective_2}
\caption{Comparison of supervised compression objectives.}
\end{table}

\begin{figure}[htb]
    \centering
	\includegraphics[width=0.4\textwidth]{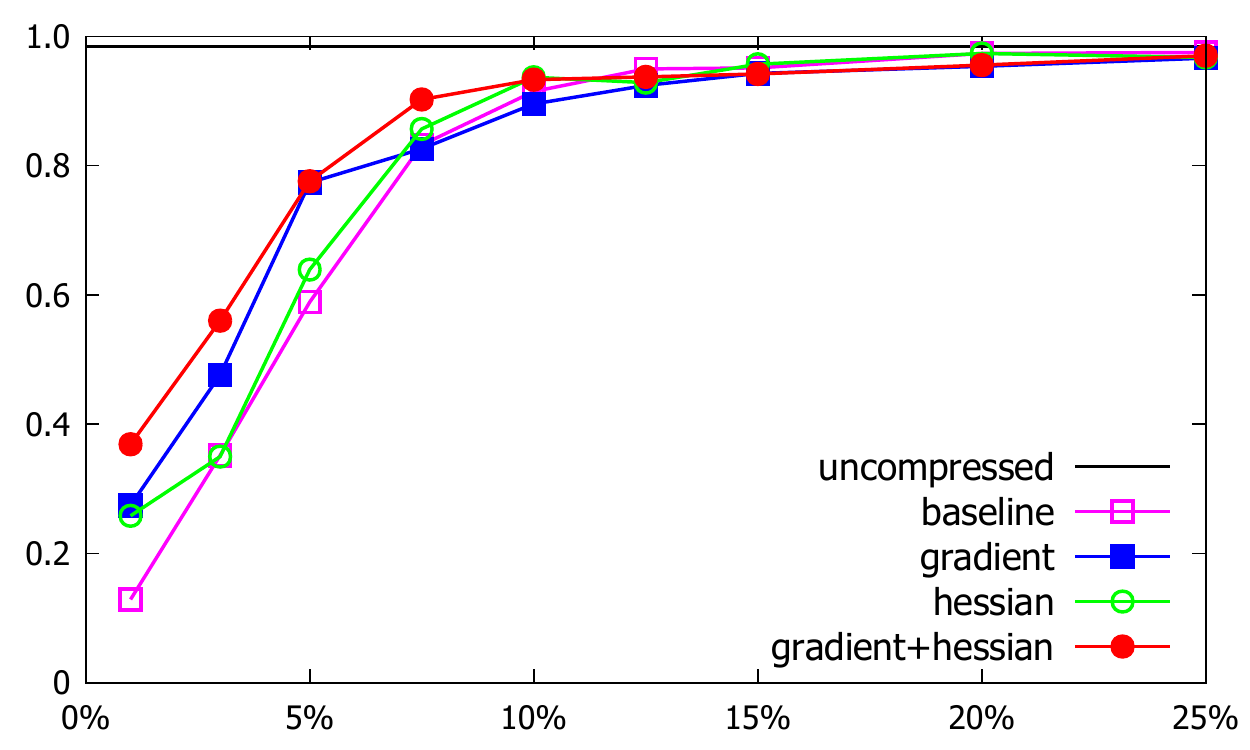}
	\put(-200,40){\rotatebox{90}{Accuracy}}
    \includegraphics[width=0.4\textwidth]{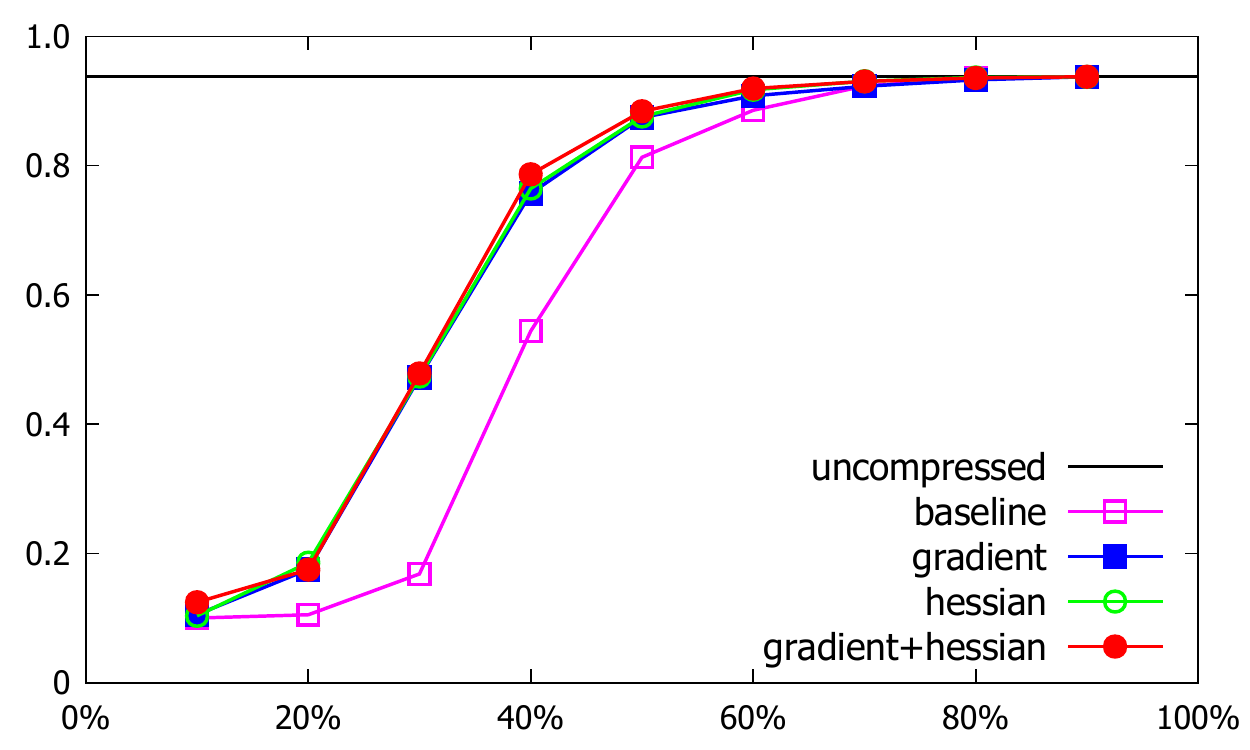}
    \includegraphics[width=0.4\textwidth]{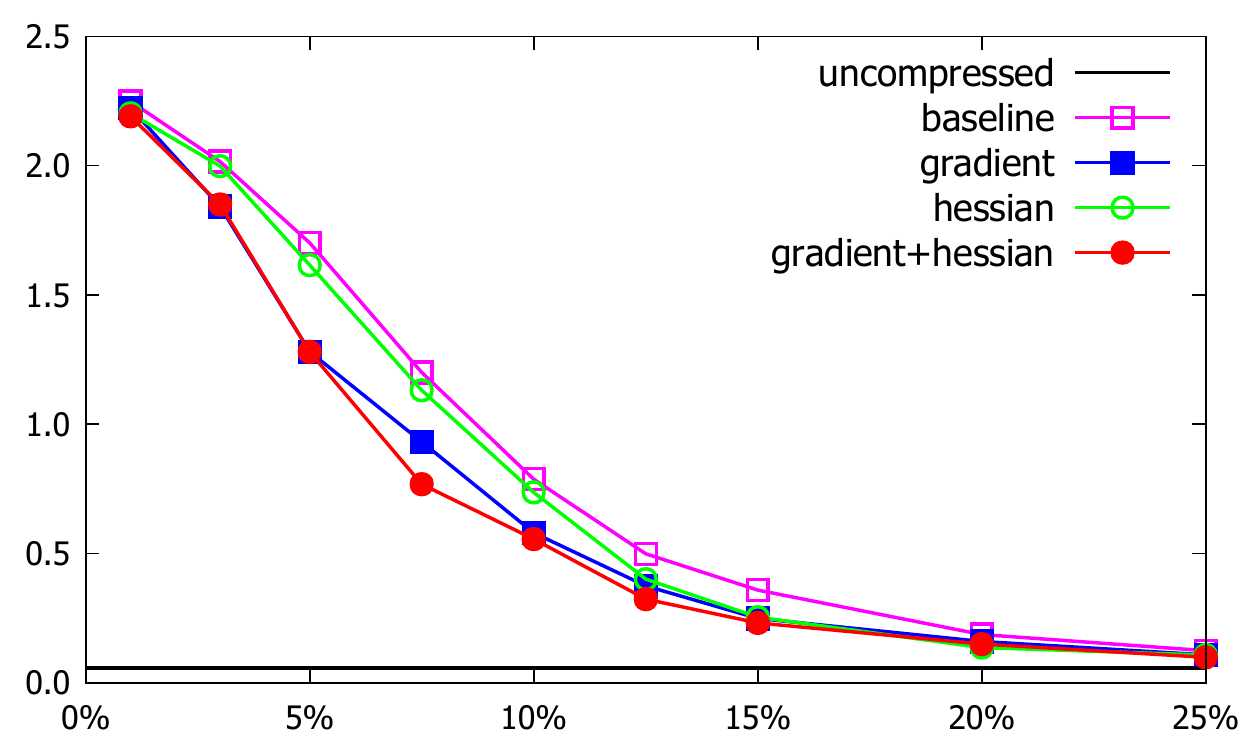}
    \put(-200,30){\rotatebox{90}{Cross Entropy}}
    \includegraphics[width=0.4\textwidth]{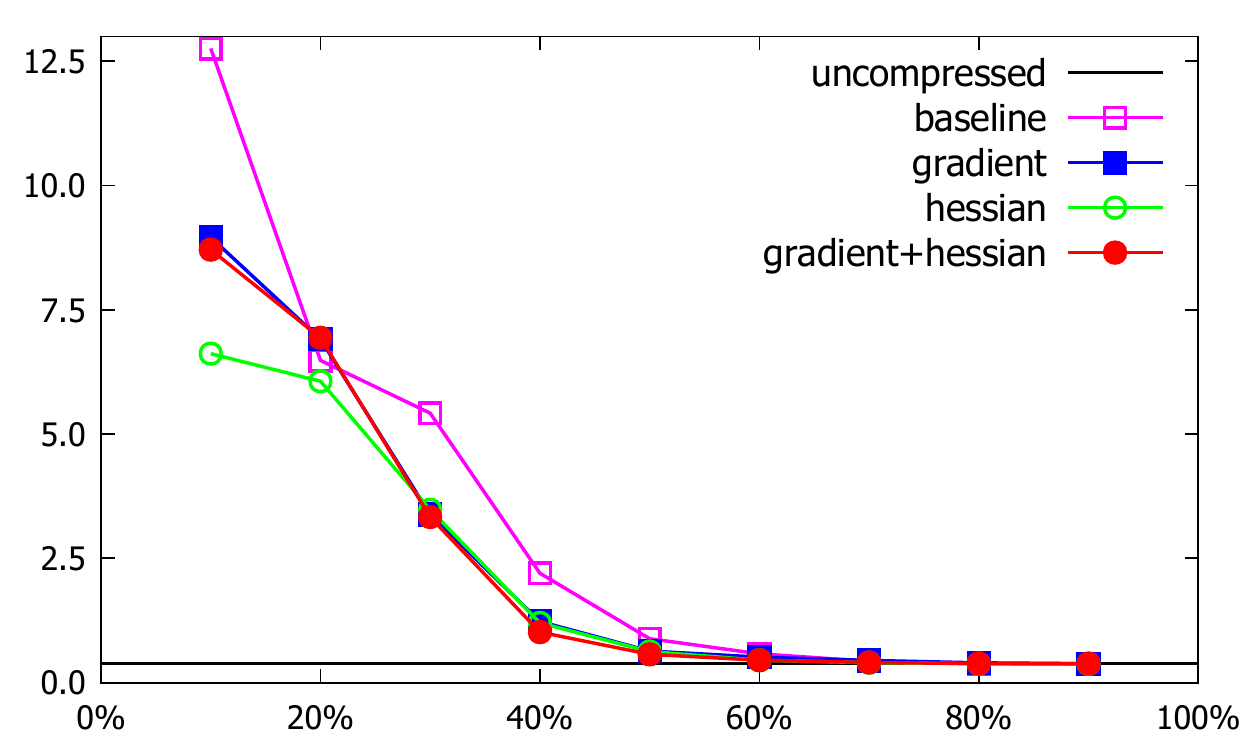}
    \put(-130,-10){Compression Ratio}
    \put(-320,-10){Compression Ratio}
    \caption{Result for supervised pruning experiment. Left: fully-connected neural network on MNIST (Top: test accuracy, Bottom: test cross entropy loss). Right: convolutional neural network on CIFAR10 (Top: test accuracy, Bottom: test cross entropy loss).}
    \label{fig:ce_pruning}
\end{figure}

\begin{figure}[htb]
    \centering
	\includegraphics[width=0.4\textwidth]{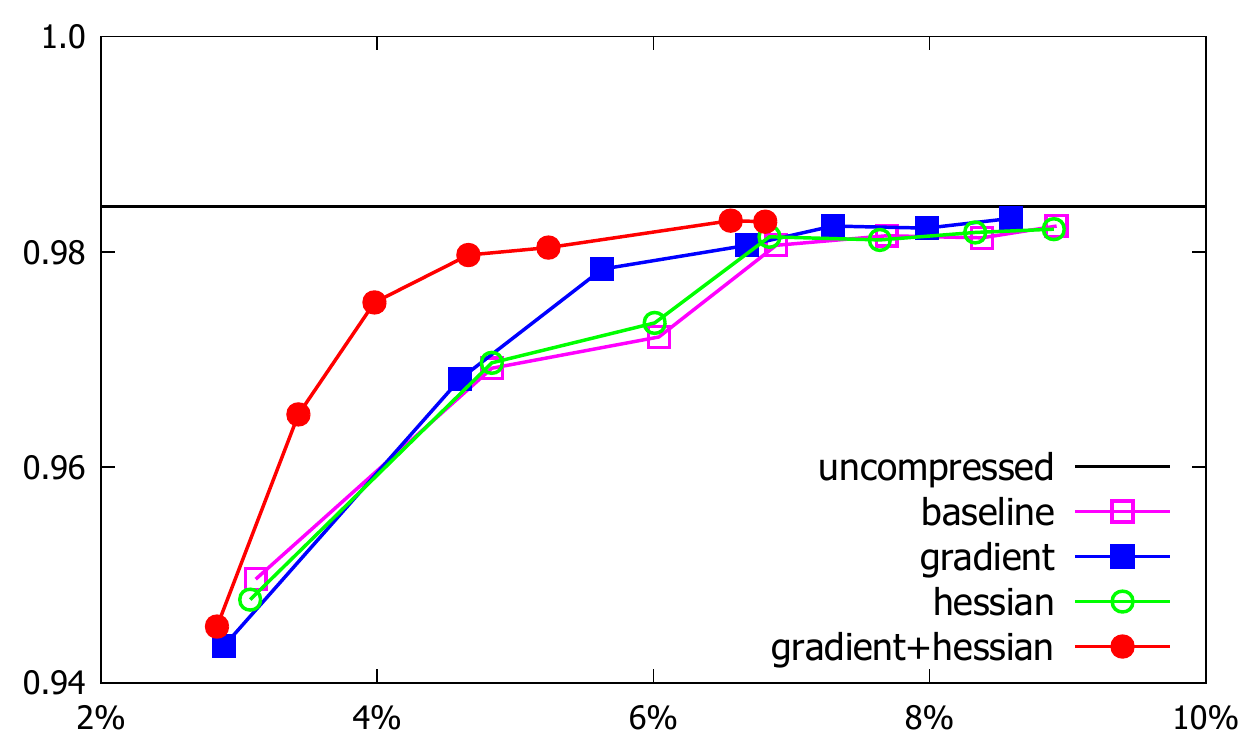}
	\put(-200,40){\rotatebox{90}{Accuracy}}
    \includegraphics[width=0.4\textwidth]{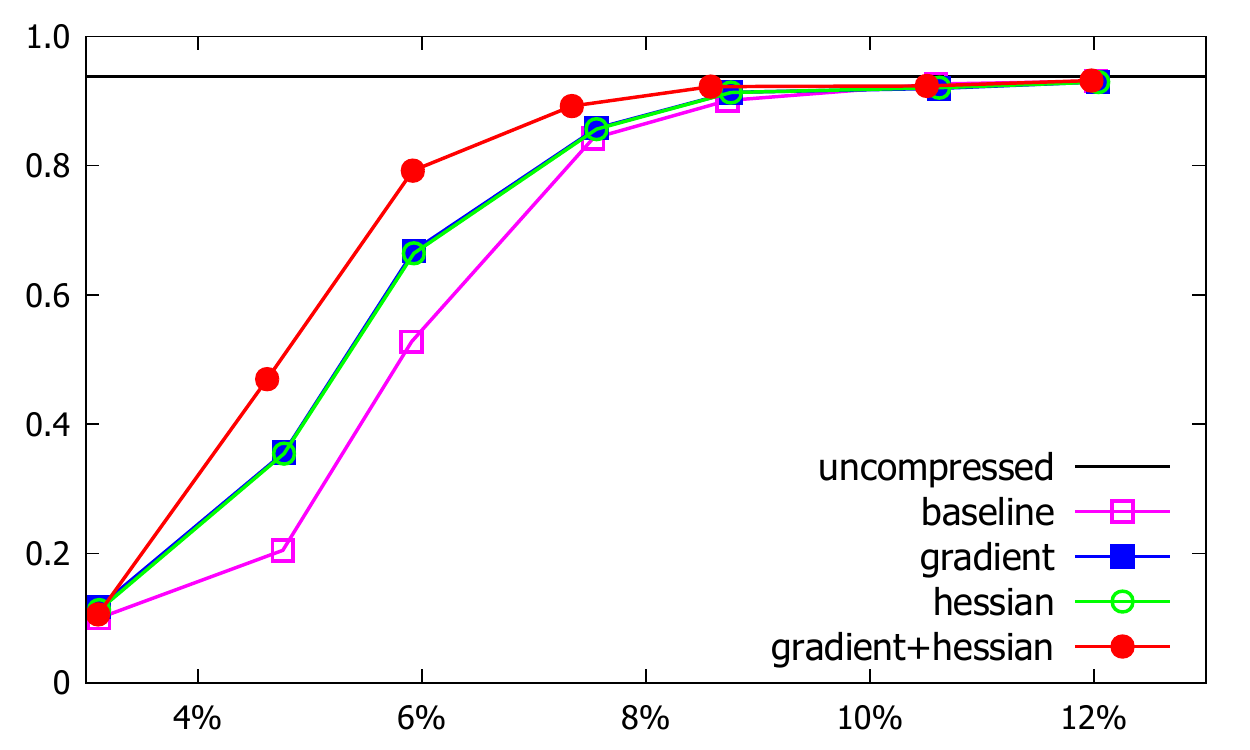}
    \includegraphics[width=0.4\textwidth]{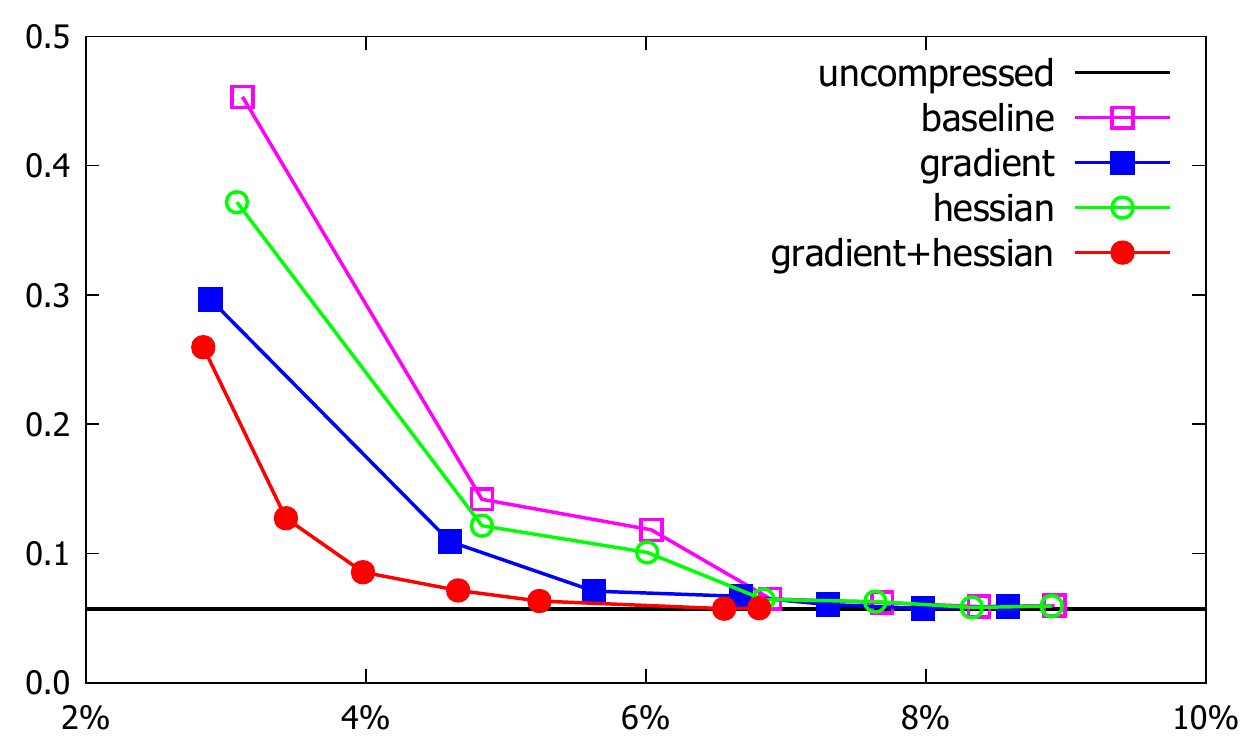}
    \put(-200,30){\rotatebox{90}{Cross Entropy}}
    \includegraphics[width=0.4\textwidth]{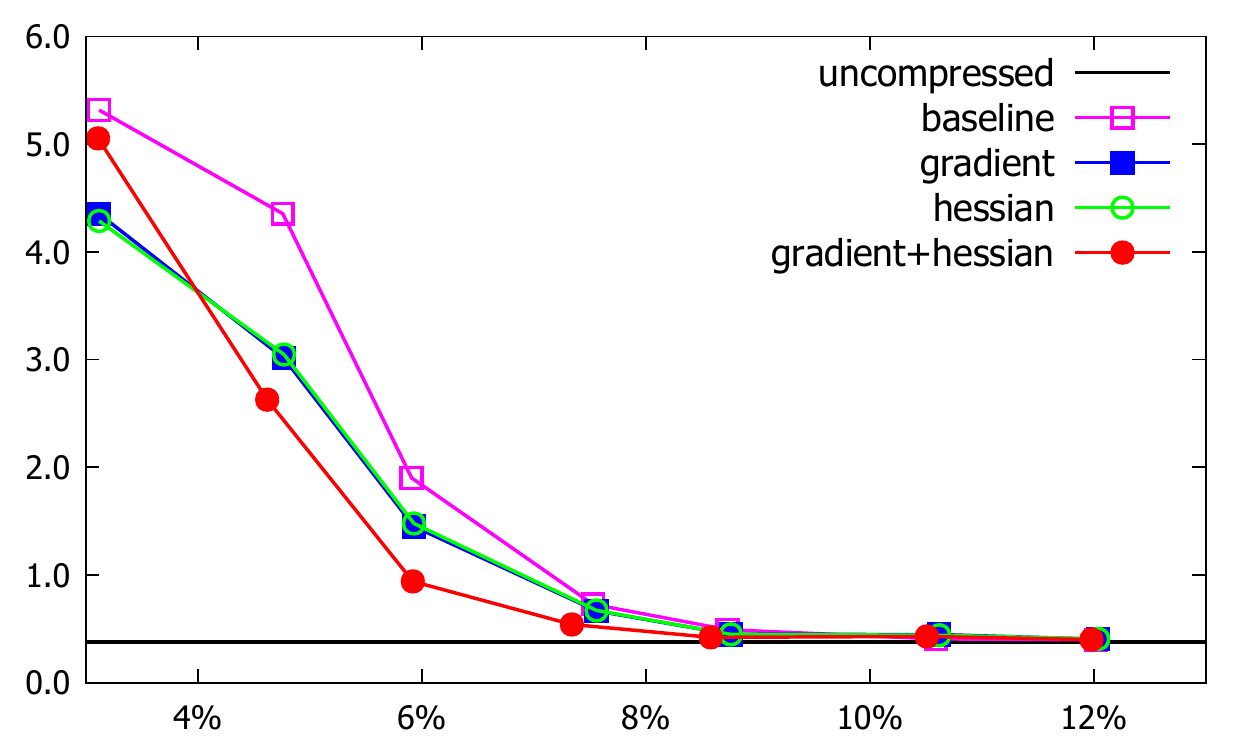}
    \put(-130,-10){Compression Ratio}
    \put(-320,-10){Compression Ratio}
    \caption{Result for supervised quantization experiment. Left: fully-connected neural network on MNIST (Top: test accuracy, Bottom: test cross entropy loss). Right: convolutional neural network on CIFAR10 (Top: test accuracy, Bottom: test cross entropy loss).}
    \label{fig:ce_quantization}
\end{figure}

 We show that results of pruning experiment in Figure~\ref{fig:ce_pruning}, and the results of quantization experiment in Figure~\ref{fig:ce_quantization}. Generally, the gradient objective and hessian objective both give better performance than baseline objective , while gradient objective is slightly than hessian objective at some points. Gradient + hessian objective gives the best overall performance. We relegate the results for CIFAR100 in Appendix~\ref{sec:append_experiment}.

{\bf Remark}. Here we define the supervised distortion function as $d(w, \hat{w}) = \mathbb{E}_{X,Y} \left[ (\mathcal{L}_w(X,Y) - \mathcal{L}_{\hat{w}}(X,Y))^2 \right]$, analogously to the distortion of regression. However, since the goal of classification is to minimize the loss function, the following definition of distortion function $\tilde{d}(w, \hat{w}) = \mathbb{E}_{X,Y} \left[ \mathcal{L}_{\hat{w}}(X,Y) - \mathcal{L}_{w}(X,Y) \right]$ is also valid and has been adopted in~\citet{lecun1990optimal} and~\citet{choi2016towards}. The main difference is --- $d(w, \hat{w})$ focus on the quality of {\em compression algorithm}, i.e., how similar is the compressed model compared to uncompressed model, whereas $\tilde{d}(w, \hat{w})$ focus on the quality of {\em compressed model}, i.e. how good is the compressed model. So $d(w, \hat{w})$ is a better criteria for the compression algorithm. Additionally, by taking second order approximation of $d(w, \hat{w})$, we have gradient+hessian objective, which shows better empirical performance than hessian objective, derived by taking second order approximation of $\tilde{d}(w, \hat{w})$. 

\section{Conclusion}

In this paper, we investigate the fundamental limit of neural network model compression algorithms. We prove a lower bound for the rate distortion function for model compression, and prove its achievability for linear model. Motivated by the rate distortion function, we propose the weight importance matrtix, and show that for one-hidden-layer ReLU network, pruning and quantization that minimizes the proposed objective is optimal. We also show the superiority of proposed objective in real neural networks. 

%\clearpage
\section*{Acknowledgement}
The authors thank Denny Zhou for initial comments and helpful discussions.

\bibliographystyle{apalike}
\bibliography{model_compression}

\clearpage
%\onecolumn
\appendix

% ----------------------------------------- 
\section{Lower bound for rate distortion function}
\label{sec:append_lower_bound}
In this section, we finish the proof of the lower bound and achievability in Section~\ref{sec:lower_bound}. Our approach is based on the water-filling approach~\cite{mcdonald1964information}.

\subsection{General lower bound}
First, we establish establishes a lower bound of the rate distortion function, which works for general models..
\begin{lemma}
The rate-distortion function $R(D) \geq \underline{R}(D) = h(W) - C$, where $C$ is the optimal value of the following optimization problem.
\begin{eqnarray*}
    \max_{P_{\hat{W}|W}} && \sum_{i=1}^m \min \Big\{ h(W_i), \frac{1}{2} \log ( 2 \pi e \mathbb{E}_{W,\hat{W}}[(W_i-\hat{W}_i)^2])\Big\} \nonumber\\
    {\rm s.t.} && E_{W, \hat{W}} \left[ d(W,\hat{W}) \right] \leq D. 
\end{eqnarray*}
\label{lem:lower_bound}
\end{lemma}
where $h(W) = - \int_{w \in \mathcal{W}} P_W(w) \log P_W(w) dw$ is the differential entropy of $W$ and $h(W_i)$ is the differential entropy of the $i$-th entry of $W$. 

\subsubsection{Proof of Lemma~\ref{lem:lower_bound}}
Recall that the rate distortion function for model compression is defined as $R(D) = \min_{P_{\hat{W}|W}: \mathbb{E}_{W, \hat{W}} [d(W, \hat{W})] \leq D} I(W; \hat{W})$. Now we lower bound the mutual information $I(W, \hat{W})$ by
\begin{eqnarray*}
    I(W;\hat{W}) &=& h(W) - h(W \,|\, \hat{W}) \nonumber,\\
    &=& h(W) - \sum_{i=1}^m h(W_i \,|\, W_1, \dots, W_{i-1},  \hat{W}_i, \dots, \hat{W}_m) \,\notag\\
    &\geq& h(W) - \sum_{i=1}^m h(W_i \,|\, \hat{W}_i ). 
\end{eqnarray*}
Here the last inequality comes from the fact that conditioning does not increase entropy. 
%Since this lower bounds holds for any unitary matrix $Q$, we can take the minimum of all unitary matrices.
%\begin{eqnarray}
%I(W;\hat{W}) &\geq& h(W) - \min_{Q \in \mathbb{U}^{m \times m}} \sum_{i=1}^m h(W_i \,|\, \hat{W}_i ).
%\end{eqnarray}
Notice that the first term $h(W)$ does not depend on the compressor. For the last term, we upper bound each term $h(W_i \,|\, \hat{W}_i )$ in two ways. On one hand, $h(W_i \,|\, \hat{W}_i )$ is upper bounded by $h(W_i)$ because conditioning does not increase entropy. On the other hand, $h(W_i \,|\, \hat{W}_i ) = h(W_i - \hat{W}_i\,|\, \hat{W}_i ) \leq h(W_i - \hat{W}_i )$, and by~\citet[Theorem 8.6.5]{cover2012elements}, differential entropy is maximized by Gaussian distribution, for given second moment. We then have:
\begin{eqnarray*}
    h(W_i \,|\, \hat{W}_i ) &\leq& \min \left\{ h(W_i), h( W_i - \hat{W}_i) \right\} \,\notag\\
    &\leq& \min \Big\{ h(W_i), \frac{1}{2} \log\left( 2 \pi e \mathbb{E}_{W,\hat{W}} [(W_i-\hat{W}_i)^2]\right)\Big\} \,\notag\\
    &=& \min \left\{ h(W_i),  \frac{1}{2} \log ( 2 \pi e \mathbb{E}_{W,\hat{W}}[(W_i-\hat{W}_i)^2]) \right\}.
\end{eqnarray*}
Therefore, the lower bound of the mutual information is given by,
\begin{eqnarray*}
&&I(W; \hat{W}) \geq h(W) - \sum_{i=1}^m \min \Big\{ h(W_i), \frac{1}{2} \log ( 2 \pi e \mathbb{E}_{W,\hat{W}}[(W_i-\hat{W}_i)^2]) \Big\}.
\end{eqnarray*}

\subsection{Lower bound for linear model}

For complex models, the general lower bound in Lemma~\ref{lem:lower_bound} is difficult to evaluate, due to the large dimension of parameters. It was shown by~\citet{jiao2017nearest} that the sample complexity to estimate differential entropy is exponential to the dimension. It's even harder to design an algorithm to achieve the lower bound. But for linear model, the lower bound can be simplified. For $f_w(x) = w^T x$, the distortion function $d(w, \hat{w})$ can be written as
\begin{eqnarray*}
d(w, \hat{w}) &=& \mathbb{E}_X \left[ (f_w(X) - f_{\hat{w}}(X) )^2 \right] = \mathbb{E}_X \left[ ( w^T X - \hat{w}^T X)^2 \right] \,\notag\\
&=& \mathbb{E}_X \left[(w-\hat{w})^T X X^T (w-\hat{w}) \right] = (w-\hat{w})^T \mathbb{E}_X [X X^T] (w-\hat{w}).
\end{eqnarray*}

Since we assumed that $\mathbb{E}[X] = 0$, $\mathbb{E}[X_i^2] = \lambda_{x, i} > 0$ and $\mathbb{E}[X_i X_j] = 0$, so the constraint in Lemma~\ref{lem:lower_bound} is given by
\begin{eqnarray*}
D &\geq& \mathbb{E}_{W, \hat{W}} \left[(W - \hat{W})^T \mathbb{E}_X [XX^T] (W-\hat{W}) \right] \,\notag\\
&=& \sum_{i=1}^m \lambda_{x,i} \underbrace{\mathbb{E}_{W,\hat{W}} \left[(W_i-\hat{W}_i)^2\right]}_{D_i}.
\end{eqnarray*}
Then the optimization problem in Lemma~\ref{lem:lower_bound} can be written as follows
\begin{eqnarray*}
\max_{p(\hat{w}|w)} && \sum_{i=1}^m \min\{ h(W_i ), \frac{1}{2} \log (2 \pi e D_i)\} \,\notag\\
{\rm s.t.} && \sum_{i=1}^{m} \lambda_{x,i} D_i \leq D.
\end{eqnarray*}
Here $W_i$ is a Gaussian random variable, so $h(W_i) = \frac{1}{2} \log (2 \pi e \mathbb{E}[W_i^2])$. The Lagrangian function of the problem is given by
\begin{eqnarray*}
&&\mathcal{L}(D_1,\dots, D_m, \mu) \,\notag\\
&=& \sum_{i=1}^m \Big( \min\{ \frac{1}{2} \log \mathbb{E}[W_i^2], \frac{1}{2} \log D_i\}  + \frac{1}{2} \log (2 \pi e) - \mu \lambda_{x,i} D_i \Big).
\end{eqnarray*}
By setting the derivative w.r.t. $D_i$ to 0, we have
\begin{eqnarray*}
0 = \frac{\partial \mathcal{L}}{\partial D_i} = \frac{1}{2D_i} - \mu \lambda_{x,i} .
\end{eqnarray*}
for all $D_i$ such that $D_i < \mathbb{E}[W_i^2]$. So the optimal $D_i$ should satisfy that $D_i \lambda_{x,i}$ is constant, for all $D_i$ such that $D_i < \mathbb{E}[W_i^2]$. Also the optimal $D_i$ is at most $\mathbb{E}[W_i^2]$. Also, since $h(W) = \frac{m}{2} \log (2 \pi e) + \frac{1}{2} \log \det (\Sigma_W)$ the lower bound is given by
\begin{eqnarray*}
		R(D) \geq \frac{1}{2} \log \det (\Sigma_W) - \sum_{i=1}^m \frac{1}{2} \log (D_i),
\end{eqnarray*}
where
\begin{eqnarray*}
	D_i = \begin{cases}
    	\mu/\lambda_{x,i} \quad {\rm if} \mu < \lambda_{x,i} \mathbb{E}_W [W_i^2] \;, \\
        \mathbb{E}_W [W_i^2]  \quad {\rm if} \mu \geq \lambda_{x,i}\mathbb{E}_W [W_i^2]  \;,
    \end{cases}
\end{eqnarray*}
where $\mu$ is chosen that $\sum_{i=1}^m \lambda_{x,i} D_i = D$.

\begin{figure}
    \centering
    \includegraphics[width=0.5\textwidth]{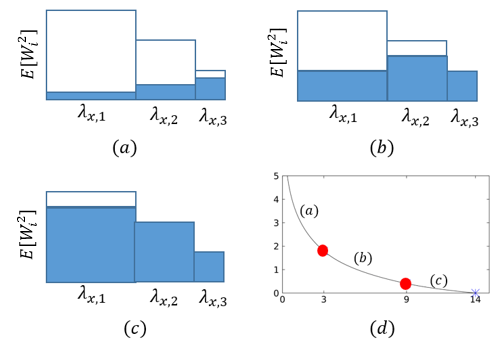}
    \caption{Illustration of ``weighted water-filling'' process.}
    \label{fig:lower_bound}
\end{figure}

This lower bound gives rise to a ``weighted water-filling'', which differs from the classical ``water-filling'' for rate-distortion of colored Gaussian source in~\citet[Figure 13.7]{cover2012elements}, since the water level's $D_i$ are proportional to $1/\lambda_{x,i}$, which is related to the input of the model rather than the parameters to be compressed. To illustrate the ``weighted water-filling'' process, we choose a simple example where $\Sigma_W = \Sigma_X = {\rm diag}[3,2,1]$. In Figure~\ref{fig:lower_bound}, the widths of each rectangle are proportional to $\lambda_{x,i}$, and the heights are proportional to $\Sigma_W = [3,2,1]$. The water level in each rectangle is $D_i$ and the volume of water is $\mu$. As $D$ starts to increase from $0$, each rectangle is filled with same volume of water ($\mu$ is the same), but the water level $D_i$'s increase with speed $1/\lambda_{x,i}$ respectively (Figure~\ref{fig:lower_bound}.(a)). This gives segment (a) of the rate distortion curve in Figure~\ref{fig:lower_bound}.(d). If $D$ is large enough such that the third rectangle is full, then $D_3$ is fixed to be $\mathbb{E}[W_3^2] = 1$, whereas $D_1$ and $D_2$ continuously increase (Figure~\ref{fig:lower_bound}.(b)). This gives segment (b) in Figure~\ref{fig:lower_bound}.(d). Keep increasing $D$ until the second rectangle is also full, then $D_2$ is fixed to be $\mathbb{E}[W_2^2] = 2$ and $D_1$ continuous increasing (Figure~\ref{fig:lower_bound} (c)). This gives segment (c) in Figure~\ref{fig:lower_bound}.(d). The entire rate-distortion function is shown in Figure~\ref{fig:lower_bound}(d), where the first red dot corresponds to the moment that the third rectangle is exactly full, and the second red dot corresponds to moment that the second rectangle is exactly full.
%\todo{Done}

\subsection{Achievability}
We prove that this lower bound is achievable. To achieve the lower bound, we construct the compression algorithm in Algorithm~\ref{algo:optimal},

\begin{algorithm}[htbp]
    \caption{Optimal compression algorithm for linear regression}
    \label{algo:optimal}
\begin{algorithmic}
    \STATE {\bf Input:} distortion $D$, covariance matrix of parameters $\Sigma_W$, covariance matrix of data $\Sigma_X = {\rm diag}[\lambda_{x,1}, \dots, \lambda_{x,m}]$. \\
    \STATE Choose $D_i$'s such that 
        \begin{eqnarray*}
            D_i = \begin{cases}
    	        \mu/\lambda_{x,i} \quad {\rm if} \mu < \lambda_{x,i} \mathbb{E}_W [W_i^2] \;, \\
                \mathbb{E}_W [W_i^2]  \quad {\rm if} \mu \geq \lambda_{x,i}\mathbb{E}_W [W_i^2]  \;,
                \end{cases}
        \end{eqnarray*} 
        where $\sum_{i=1}^m \lambda_{x,i} D_i = D$.
    \FOR{$i=1$ to $m$}
        \IF{$D_i = \mu/\lambda_{x,i}$}
            \STATE Choose $\hat{W}_i = 0$
        \ELSE
            \STATE Choose a conditional distribution $P_{\hat{W}_i|W_i}$ such that $W_i = \hat{W}+Z_i$ where  $Z_i \sim \mathcal{N}(0, D_i)$, $\hat{W}_i \sim \mathcal{N}(0, \mathbb{E}_W [W_i^2] - D_i)$ and $\hat{W}_i$ is independent of $Z_i$.
        \ENDIF
    \ENDFOR
    \STATE Combine the conditional probability distributions by $P_{\hat{W}|W} = \prod_{i=1}^m P_{\hat{W}_i|W_i}$.
\end{algorithmic}
\end{algorithm}

Intuitively, the optimal compressor does the following: (1) Find the optimal water levels $D_i$ for ``weighted water filling''. (2) For the entries where the corresponding rectangles are full, simply discard the entries; (3) for the entries where the corresponding rectangles are not full, add a noise which is independent of $\hat{W}_i$ and has a variance proportional to the water level. That is possible since $W$ is Gaussian. (4) Combine the conditional probabilities. 

To see that this compressor is optimal, we will check that the compressor makes all the inequalities become equality. Here is all the inequalities used in the proof.
\begin{itemize}
    \item $h(W_i \,|\, W_1, \dots, W_{i-1},  \hat{W}_i, \dots, \hat{W}_m) \leq h(W_i | \hat{W}_i)$ for all $i=1 ... m$. It becomes equality by $P_{\hat{W}|W} = \prod_{i=1}^m P_{\hat{W}_i|W}$.
    \item Either
        \begin{itemize}
            \item $h(W_i | \hat{W}_i) \leq h(W_i)$. It becomes equality for those $\hat{W}_i = 0$.
            \item $h(W_i - \hat{W}_i | \hat{W}_i) \leq h(W_i - \hat{W}_i) \leq \frac{1}{2} \log (2 \pi e \mathbb{E}_{W, \hat{W}} [(W_i - \hat{W})^2])$. It becomes equality for those $\hat{W}_i$'s such that $W_i - \hat{W}_i$ is independent of $\hat{W}_i$ and $W_i - \hat{W}_i$ is Gaussian.
        \end{itemize}
    \item The ``water levels'' $D_i$. It becomes equality by choosing the $D_i$'s according to Lagrangian conditions.
\end{itemize}
Therefore, Algorithm~\ref{algo:optimal} gives a compressor $P^{(D)}_{\hat{W}|W}$ such that $\mathbb{E}_{P_W \circ P^{(D)}_{\hat{W}|W}} [d(W, \hat{W})] = D$ and $I(W; \hat{W}) = \underline{R}(D)$, hence the lower bound is tight.

\section{Proof of Theorem~\ref{thm:optimal_relu}}
\label{sec:append_proof}
In this section, we provide the proof of Theorem~\ref{thm:optimal_relu}. For simplicity let $\sigma(t) = t \mathbb{I}\{t \geq 0\}$ denotes the ReLU activation function. First we deal with the objective of the compression algorithm,
\begin{eqnarray*}
    (w - \hat{w})^T I_w (w - \hat{w}) &=& (w - \hat{w})^T \mathbb{E}_X \left[ \nabla_w f_w(x) \nabla_w f_w(x)^T \right] (w - \hat{w}) \,\notag\\
    &=& (w - \hat{w})^T \mathbb{E}_X \left[ \nabla_w \sigma(w^T x) \nabla_w \sigma(w^T x)^T \right] (w - \hat{w})\,\notag\\
    &=& (w - \hat{w})^T \mathbb{E}_X \left[ x^T (\sigma'(w^T x))^2 x \right] (w - \hat{w}) \,\notag\\
    &=& \mathbb{E}_X \left[ \mathbb{I}\{w^T x \geq 0\} ((w - \hat{w})^T x)^2 \right]
\end{eqnarray*}
Notice that $x$ is jointly Gaussian random variable with zero mean and non-degenerate variance, so the distribution of $x$ is equivalent to the distribution of $-x$. Therefore,
\begin{eqnarray*}
    &&\mathbb{E}_X [\mathbb{I}\{w^T x \geq 0\} ((w - \hat{w})^T x)^2] = \int_{x: w^T x \geq 0} ((w - \hat{w}^T) x)^2 dx \,\notag\\
    &=& \frac{1}{2} \left( \int_{x: w^T x \geq 0} ((w - \hat{w}^T) x)^2 dx + \int_{x: w^T x \leq 0} ((w - \hat{w}^T) x)^2 dx \right) \,\notag\\
    &=& \frac{1}{2} \int_{x \in \mathbb{R}}^d ((w - \hat{w}^T) x)^2 dx = \frac{1}{2} (w-\hat{w})^T \Sigma_X (w-\hat{w})
\end{eqnarray*}

So minimizing the gradient-squared based loss is equivalent to minimizing $(w-\hat{w})^T \Sigma_X (w-\hat{w})$. Similarly, the condition $\hat{w} I_w (w - \hat{w}) = 0$ is equivalent to $\hat{w} \Sigma_X (w - \hat{w}) = 0$. Now we deal with the MSE loss function $\mathbb{E}[ (f_w(x) - f_{\hat{w}}(x))^2]$. We utilize the Hermite polynomials and Fourier analysis on Gaussian space. We use the following key lemma,
\begin{lemma}(\citet[Claim 4.3]{ge2017learning})
    Let $f$, $g$ be two functions from $\mathbb{R}$ to $\mathbb{R}$ such that $f^2, g^2 \in L^2(\mathbb{R}, e^{-x^2/2})$. The for any unit vectors $u, v$, we have that
    \begin{eqnarray*}
        \mathbb{E}_{x \in \mathcal{N}(0, I_{d\times d})} [f(u^T x)g(v^T x)] = \sum_{p=0}^{\infty} \hat{f}_p \hat{g}_p (u^T v)^p
    \end{eqnarray*}
    where $\hat{f}_p = \mathbb{E}_{x \in \mathcal{N}(0,1)} [f(x) h_p(x)]$ is the $p$-th order coefficient of $f$, where $h_p$ is the $p$-th order probabilists' Hermite polynomial. 
\end{lemma}

Please see Section 4.1 in~\citet{ge2017learning} for more backgrounds of the Hermite polynomials and Fourier analysis on Gaussian space. For ReLU function, the coefficients are given by $\hat{\sigma}_0 = \frac{1}{\sqrt{2 \pi}}$, $\hat{\sigma}_1 = \frac{1}{2}$. For $p \geq 2$ and even, $\hat{\sigma}_p = \frac{((p-3)!!)^2}{\sqrt{2 \pi p!}}$. For $p \geq 2$ and odd, $\hat{\sigma}_p = 0$. Since $X \sim \mathcal{N}(0, \Sigma_X)$, we can write $x = \Sigma_X^{1/2} z$, where $z \sim \mathcal{N}(0, I_d)$. So for any compressed weight $\hat{w}$, we have
\begin{eqnarray*}
    &&\mathbb{E}_X \left[ (f_w(x) - f_{\hat{w}}(x))^2 \right] = \mathbb{E}_X \left[ (\sigma(w^T x) - \sigma(\hat{w}^T x))^2 \right] \,\notag\\
    &=& \mathbb{E}_{z \in \mathcal{N}(0, I_d)} [(\sigma(w^T \Sigma_X^{1/2} z) - \sigma(\hat{w}^T \Sigma_X^{1/2} z))^2] \,\notag\\
    &=& \mathbb{E}_{z \in \mathcal{N}(0, I_d)} [\sigma(w^T \Sigma_X^{1/2} z)^2] - 2 \mathbb{E}_{z \in \mathcal{N}(0, I_d)} [ \sigma(w^T \Sigma_X^{1/2} z) \sigma(\hat{w}^T \Sigma_X^{1/2} z)] + \mathbb{E}_{z \in \mathcal{N}(0, I_d)} [\sigma(\hat{w}^T \Sigma_X^{1/2} z)^2] \,\notag\\
    &=& \sum_{p=0}^{\infty} \hat{\sigma}_p^2 (w^T \Sigma_X w)^p - 2 \sum_{p=0}^{\infty} \hat{\sigma}_p^2 (w^T \Sigma_X \hat{w})^p + \sum_{p=0}^{\infty} \hat{\sigma_p}^2 (\hat{w}^T \Sigma_X \hat{w})^p \,\notag\\
    &=& \sum_{p=0}^{\infty} \hat{\sigma}_p^2 \left( \underbrace{(w^T \Sigma_X w)^p - 2 (w^T \Sigma_X \hat{w})^p + (\hat{w}^T \Sigma_X \hat{w})^p}_{D_p(w, \hat{w})} \right)
\end{eqnarray*}

Now we can see that $D_0(w, \hat{w}) = 0$. $D_1(w, \hat{w}) = w^T \Sigma_X w - 2 w^T \Sigma_X \hat{w} + \hat{w}^T \Sigma_X w = (w-\hat{w})^T \Sigma_X (w-\hat{w})$, is just the objective. The following lemma gives the minimizer of $D_p(w, \hat{w})$ for higher order $p$.

\begin{lemma}
If $\hat{w}^*$ satisfies $\hat{w}^* \Sigma_X (\hat{w} - w) = 0$ and 
\begin{eqnarray*}
    \hat{w}^* = \arg\min_{\hat{s} \in \mathcal{W}} D_1(w, \hat{w})
\end{eqnarray*}
for some constrained set $\mathcal{W}$. Then for any $p \geq 2$ and even, we have
\begin{eqnarray*}
    \hat{w}^* = \arg\min_{\hat{w} \in \mathcal{W}} D_p(w, \hat{w})
\end{eqnarray*}
\label{lem:d_p}
\end{lemma}

Since the coefficients $\hat{\sigma}_p$ is zero for $p \geq 3$ and odd, so if a compressed weight $\hat{w}$ satisfied $\hat{w} \Sigma_X (\hat{w}-w) = 0$ and minimizes $D_1(\hat{w}, w) = (\hat{w} - w)^T \Sigma_X (\hat{w}-w)$, then it is the minimizer for all $D_p(w, \hat{w})$ for even $p$, therefore a minimizer of the MSE loss.

\subsection{Proof of Lemma~\ref{lem:d_p}}
For simplicity of notation, define $A = w^T \Sigma_X w$, $B = \hat{w}^T \Sigma_X (\hat{w} - w)$ and $C = D_1(w, \hat{w}) =  (\hat{w} - w)^T \Sigma_X (\hat{w} - w)$. For all compressors, we have $C \leq A$. Therefore, $w^T \Sigma_X \hat{w} = A + B - C$ and $\hat{w}^T \Sigma_X \hat{w} = A + 2B - C$. So
\begin{eqnarray*}
    D_p(w, \hat{w}) &=& A^p - 2 (A + B - C)^p + (A + 2B - C)^p
\end{eqnarray*}
First notice that 
\begin{eqnarray*}
\frac{\partial D_{p}(w, \hat{w})}{\partial B} = 2p ((A+2B-C)^{p-1} - (A+B-C)^{p-1}).
\end{eqnarray*} 
For even $p \geq 2$, $x^{p-1}$ is monotonically increasing, so $(A+2B-C)^{p-1} > (A+B-C)^{p-1}$ if $B > 0$ and vice versa. Therefore, for fixed $A$ and $C$, $D_p(w, \hat{w})$ is monotonically increasing for positive $B$ and decreasing for negative $B$. Therefore, $D_p(w, \hat{w})$ is minimized when $B = 0$, and the minimal value is $D_p(w, \hat{w}) = A^p - 2(A-C)^p + (A-C)^p = A^p - (A-C)^p$, which is monotonically increasing with respect to $C$. So if $\hat{w}^*$ satisfies $B=0$ and is a minimzer of $C = D_1(w, \hat{w})$, it is also a minimizer for $D_p(w, \hat{w})$ for all $p \geq 2$ and even.

\section{Details of the experiments}
\label{sec:append_experiment}

In this appendix, we give some details of the experiment and additional experiments which are omitted in the main text.

\subsection{Additional experiment results}

We present the experiment results for CIFAR100 here, due to page limit of the main text.

\begin{figure}[htbp]
    \centering
	\includegraphics[width=0.3\textwidth]{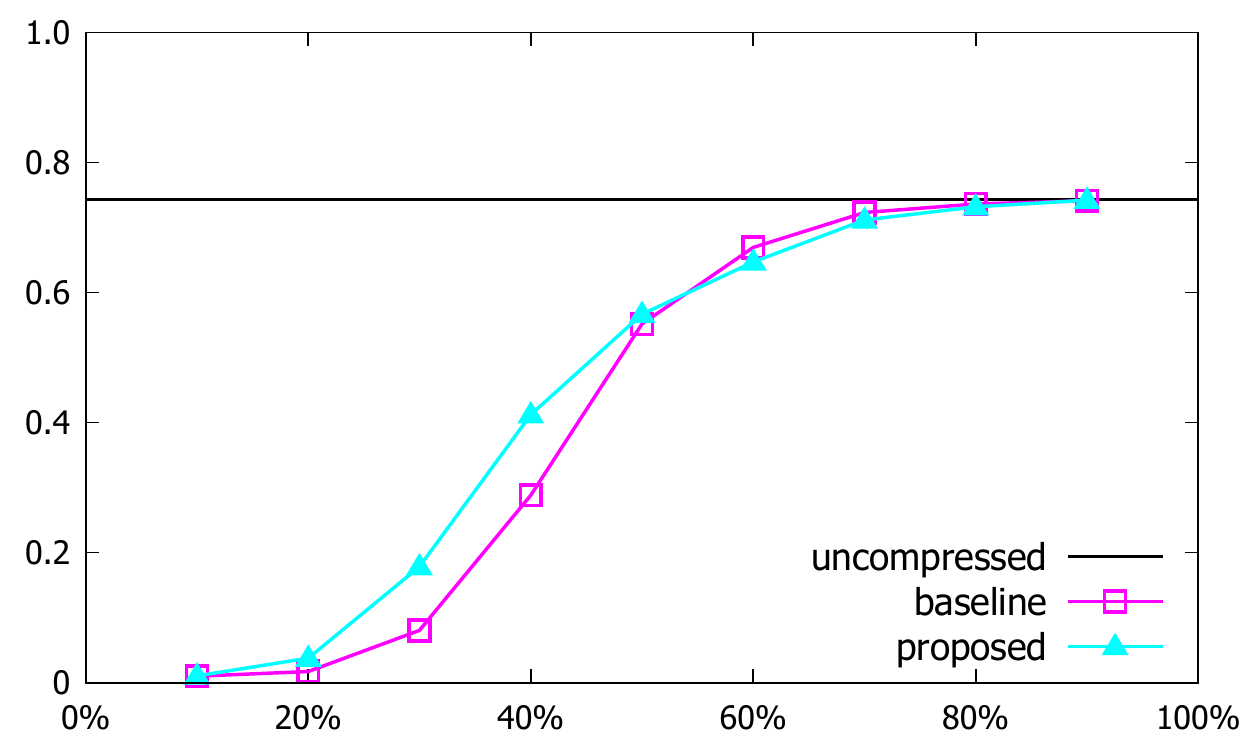}
	\put(-150,25){\rotatebox{90}{Accuracy}}
    \includegraphics[width=0.3\textwidth]{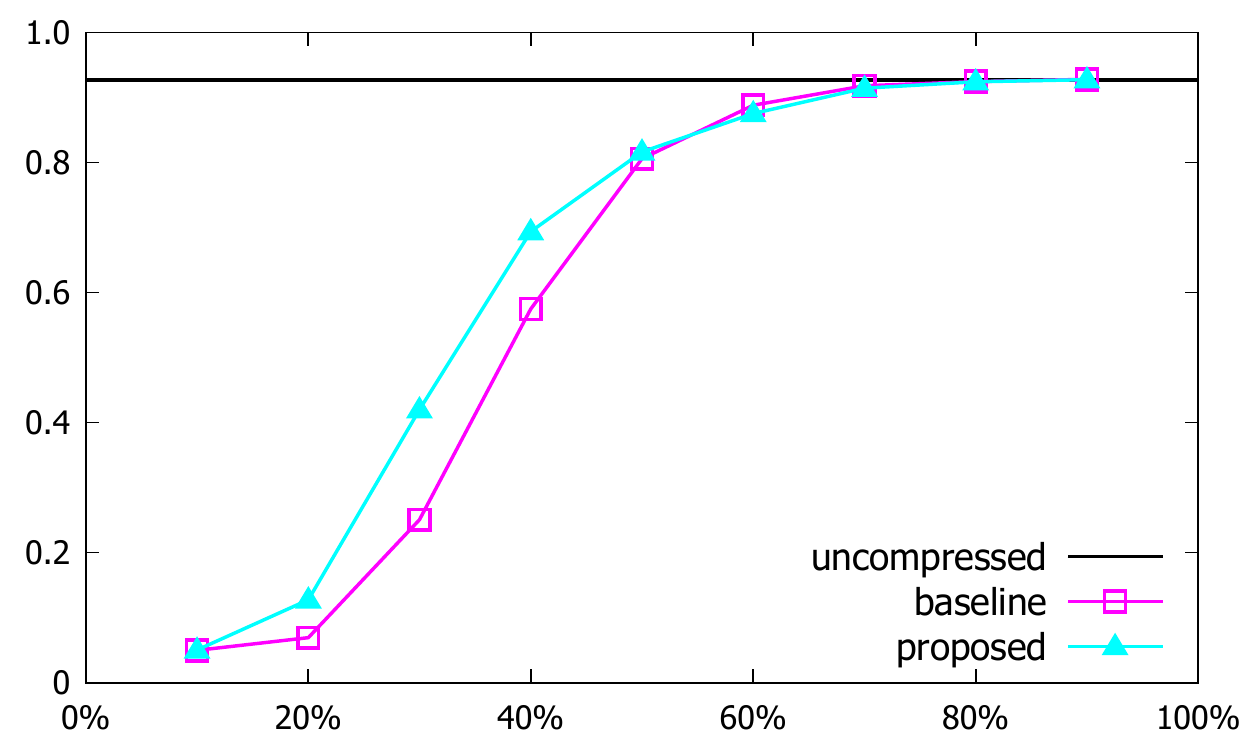}
    \vspace{0.2in}
    \includegraphics[width=0.3\textwidth]{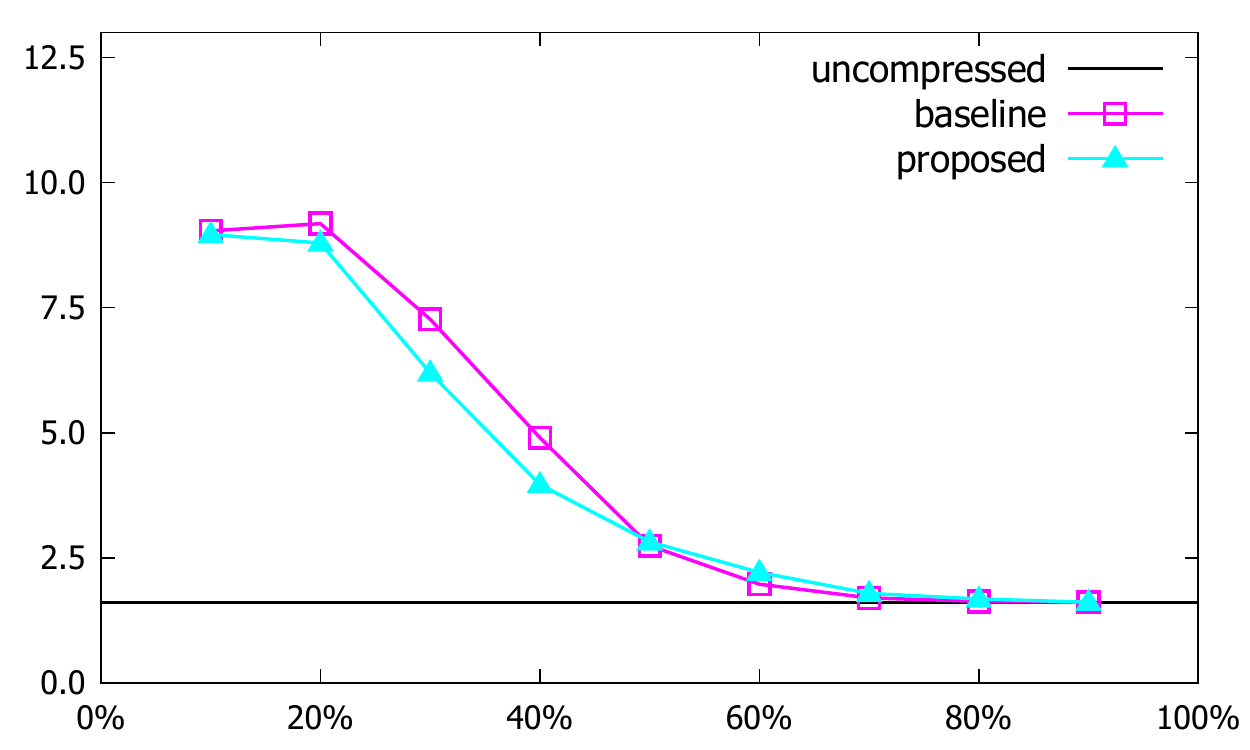}
    \put(-148,15){\rotatebox{90}{Cross Entropy}}
    \put(-110,-10){Compression Ratio}
    \put(-260,-10){Compression Ratio}
    \put(-400,-10){Compression Ratio}
    \caption{Result for unsupervised pruning experiment for CIFAR 100 experiment. Left: top-1 accuracy. Middle: top-5 accuracy. Right: cross entropy loss.}
    \label{fig:KL_pruning_100}
\end{figure}

\begin{figure}[htbp]
    \centering
	\includegraphics[width=0.3\textwidth]{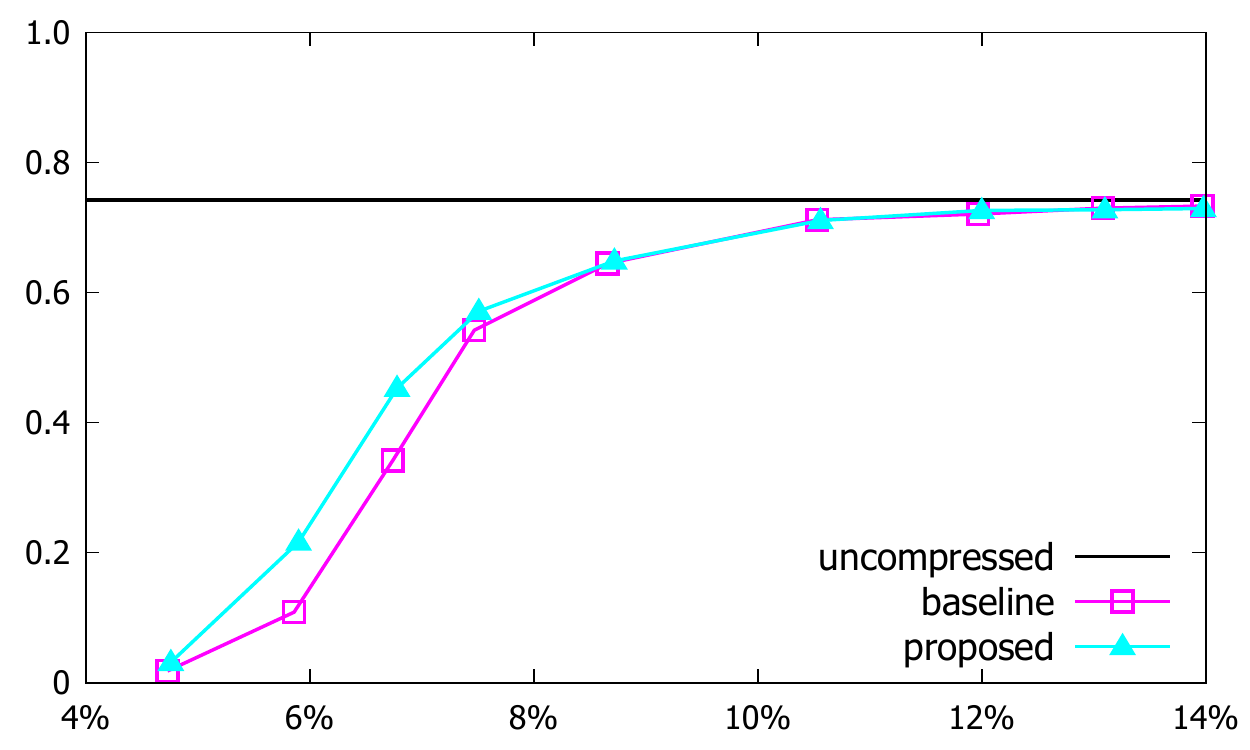}
	\put(-150,25){\rotatebox{90}{Accuracy}}
    \includegraphics[width=0.3\textwidth]{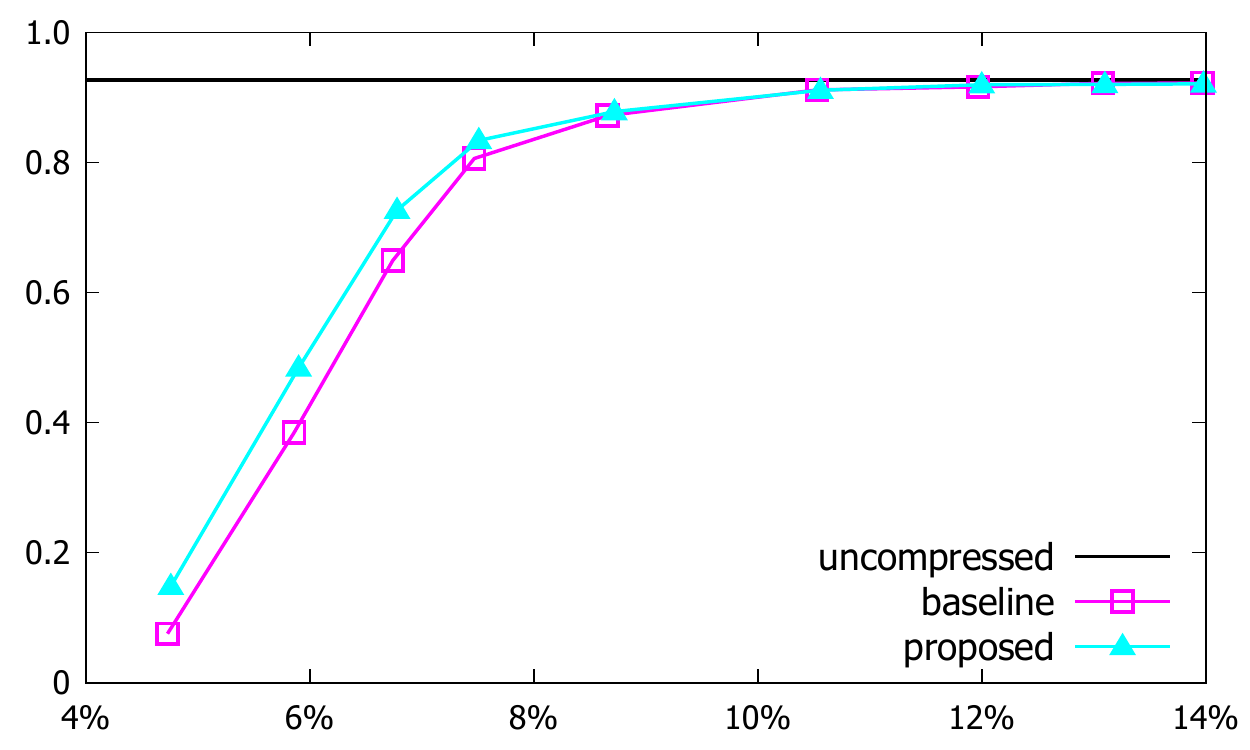}
    \vspace{0.2in}
    \includegraphics[width=0.3\textwidth]{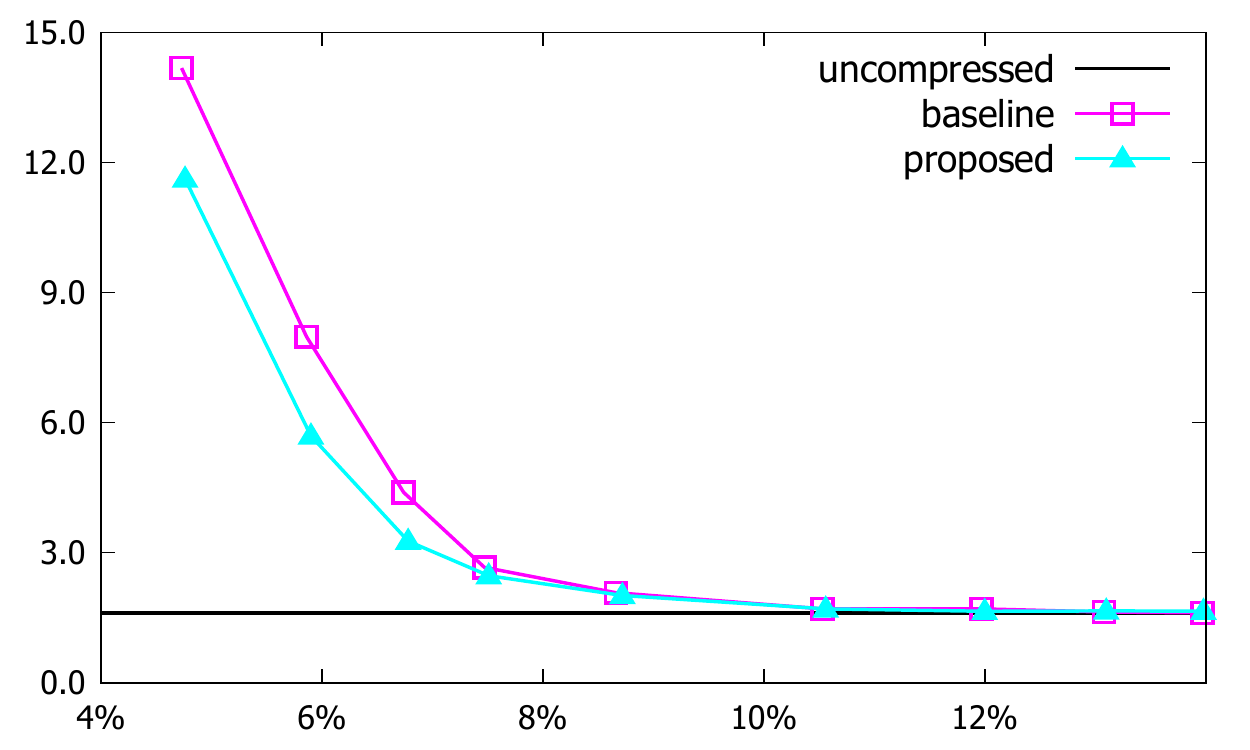}
    \put(-148,15){\rotatebox{90}{Cross Entropy}}
    \put(-110,-10){Compression Ratio}
    \put(-260,-10){Compression Ratio}
    \put(-400,-10){Compression Ratio}
    \caption{Result for unsupervised quantization experiment for CIFAR 100 experiment. Left: top-1 accuracy. Middle: top-5 accuracy. Right: cross entropy loss.}
    \label{fig:KL_quantization_100}
\end{figure}

In Figure~\ref{fig:KL_pruning_100} and Figure~\ref{fig:KL_quantization_100}, we show the result for unsupervised pruning and quantization, introduced in Section~\ref{sec:experiment}.1. We can see that, similar to the experiments of MNIST and CIFAR10, the proposed objectives gives better accuracy and smaller loss than the baseline.

\begin{figure}[htbp]
    \centering
	\includegraphics[width=0.3\textwidth]{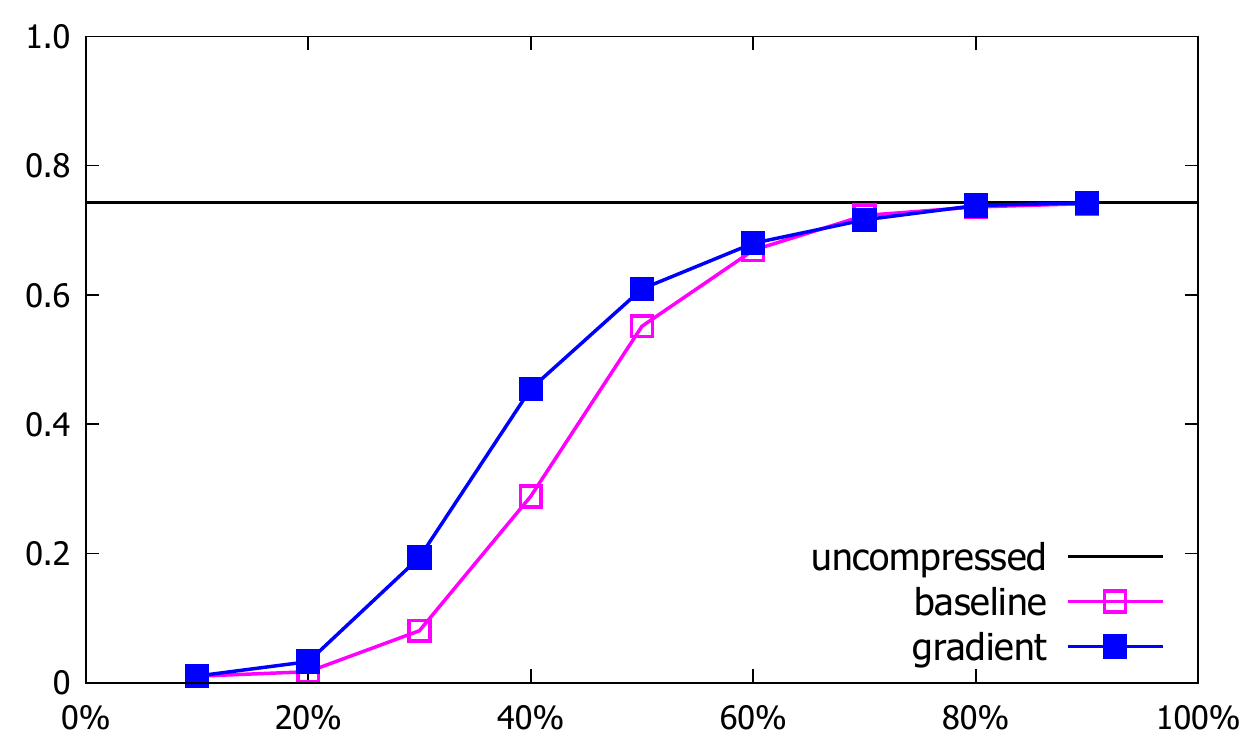}
	\put(-150,25){\rotatebox{90}{Accuracy}}
    \includegraphics[width=0.3\textwidth]{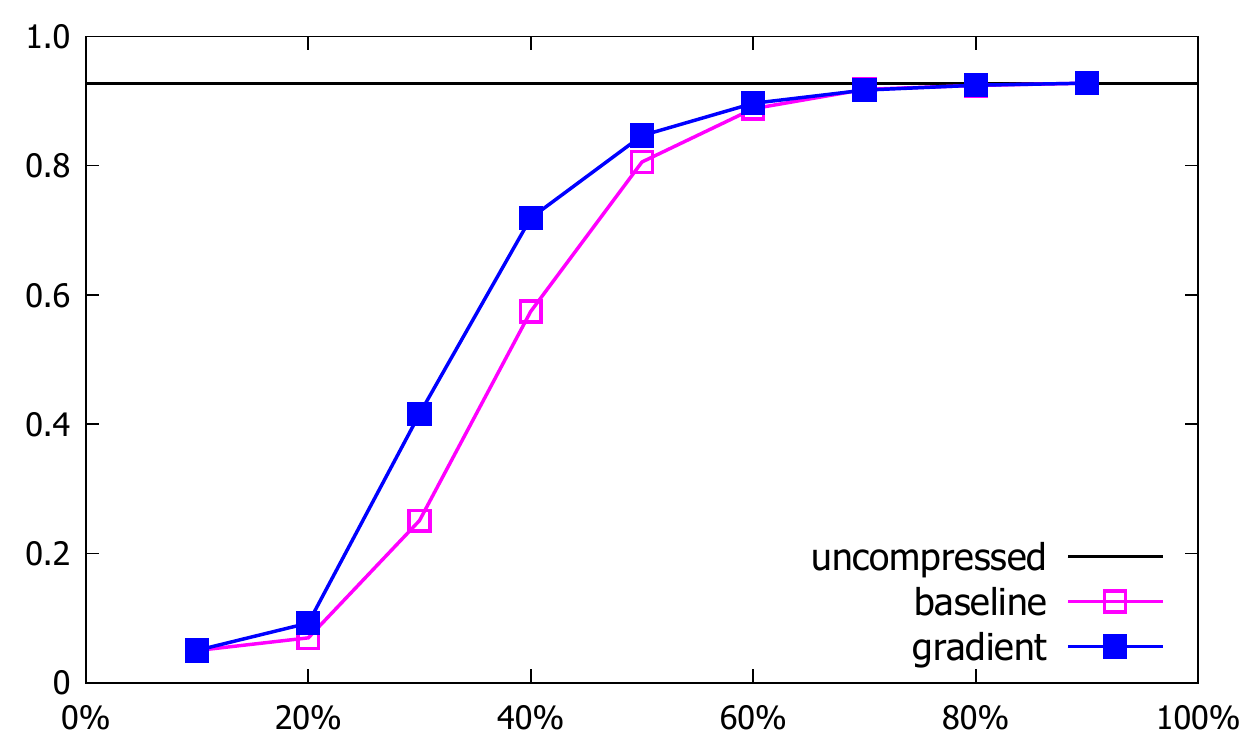}
    \vspace{0.2in}
    \includegraphics[width=0.3\textwidth]{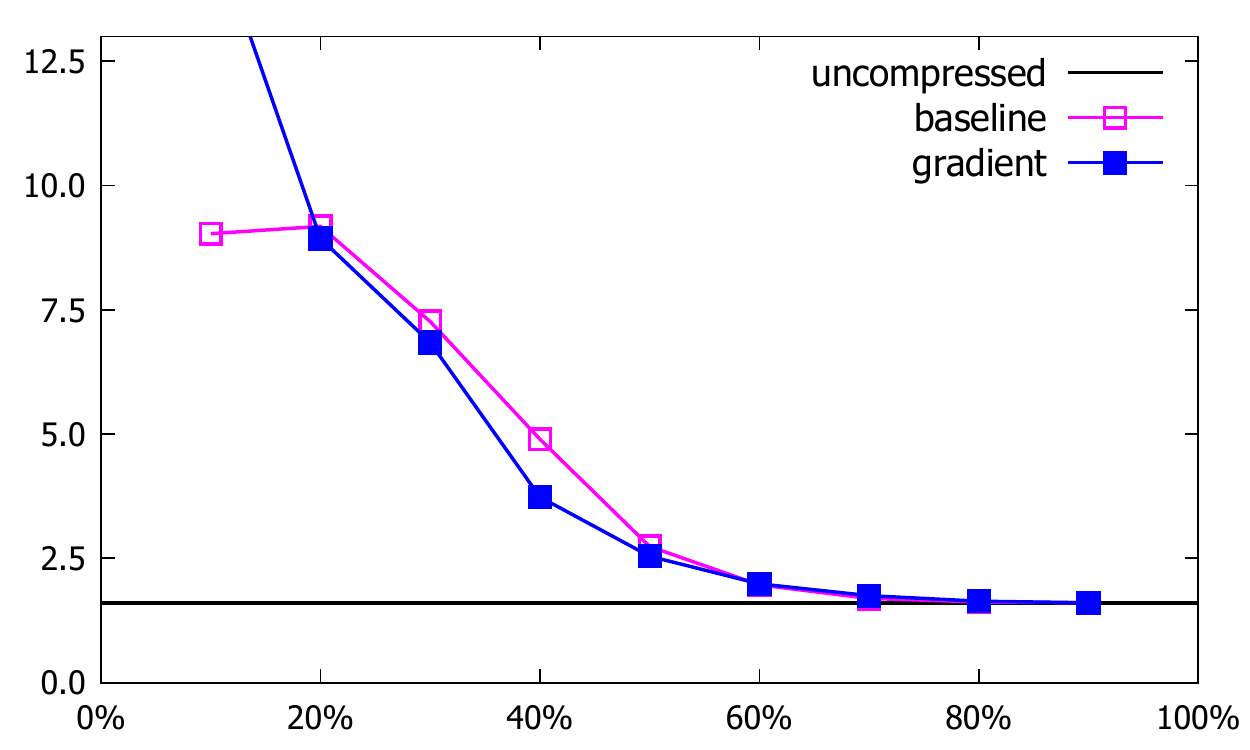}
    \put(-148,15){\rotatebox{90}{Cross Entropy}}
    \put(-110,-10){Compression Ratio}
    \put(-260,-10){Compression Ratio}
    \put(-400,-10){Compression Ratio}
    \caption{Result for supervised pruning experiment for CIFAR 100 experiment. Left: top-1 accuracy. Middle: top-5 accuracy. Right: cross entropy loss.}
    \label{fig:ce_pruning_100}
\end{figure}

\begin{figure}[htbp]
    \centering
	\includegraphics[width=0.3\textwidth]{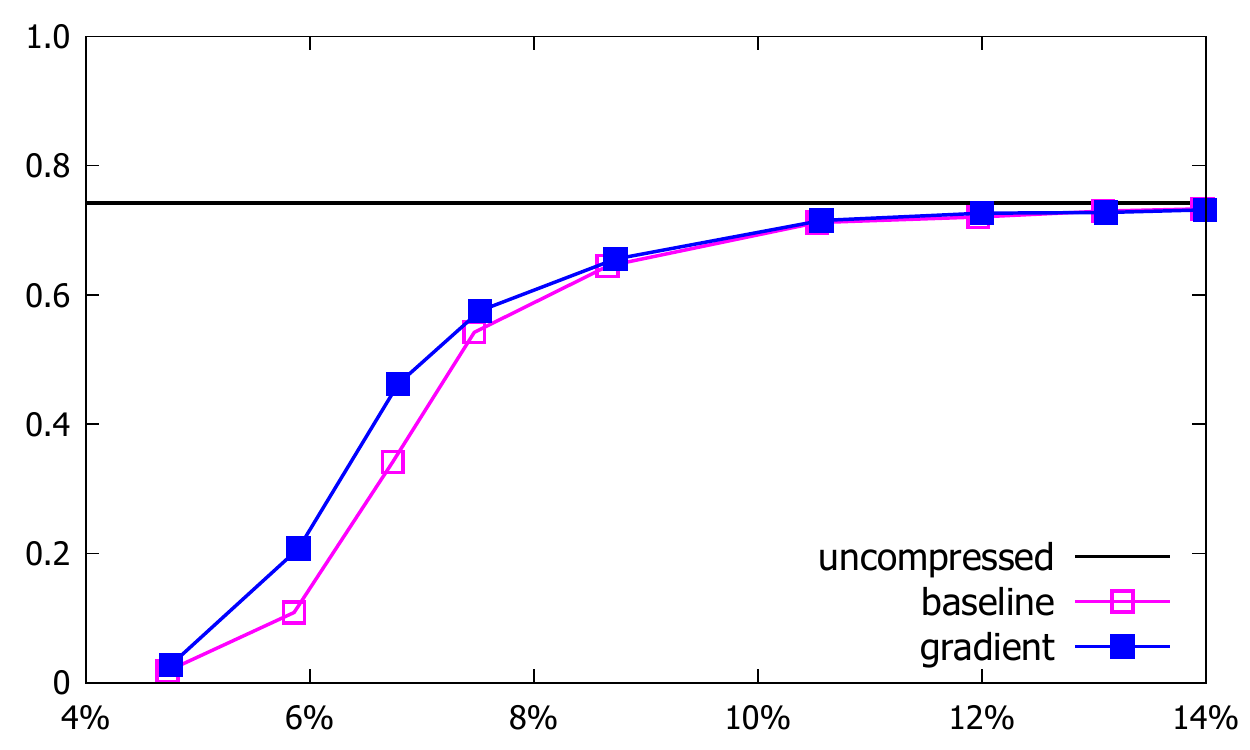}
	\put(-150,25){\rotatebox{90}{Accuracy}}
    \includegraphics[width=0.3\textwidth]{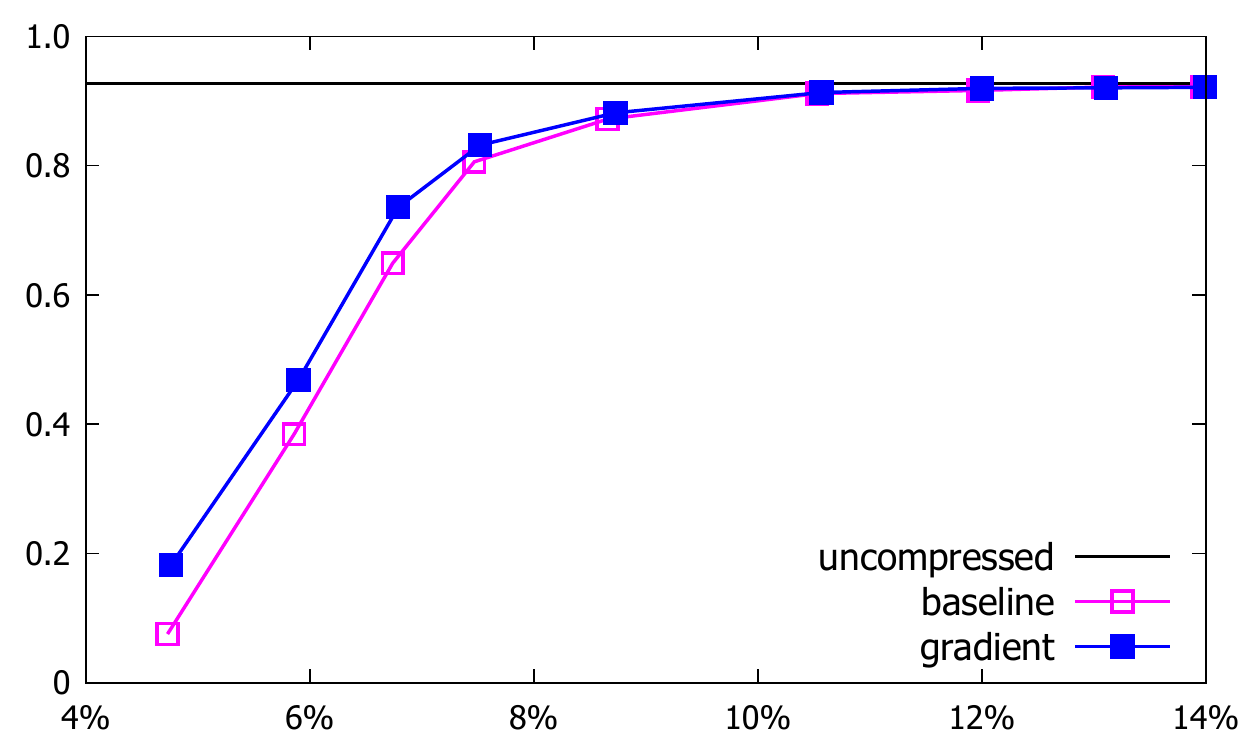}
    \vspace{0.2in}
    \includegraphics[width=0.3\textwidth]{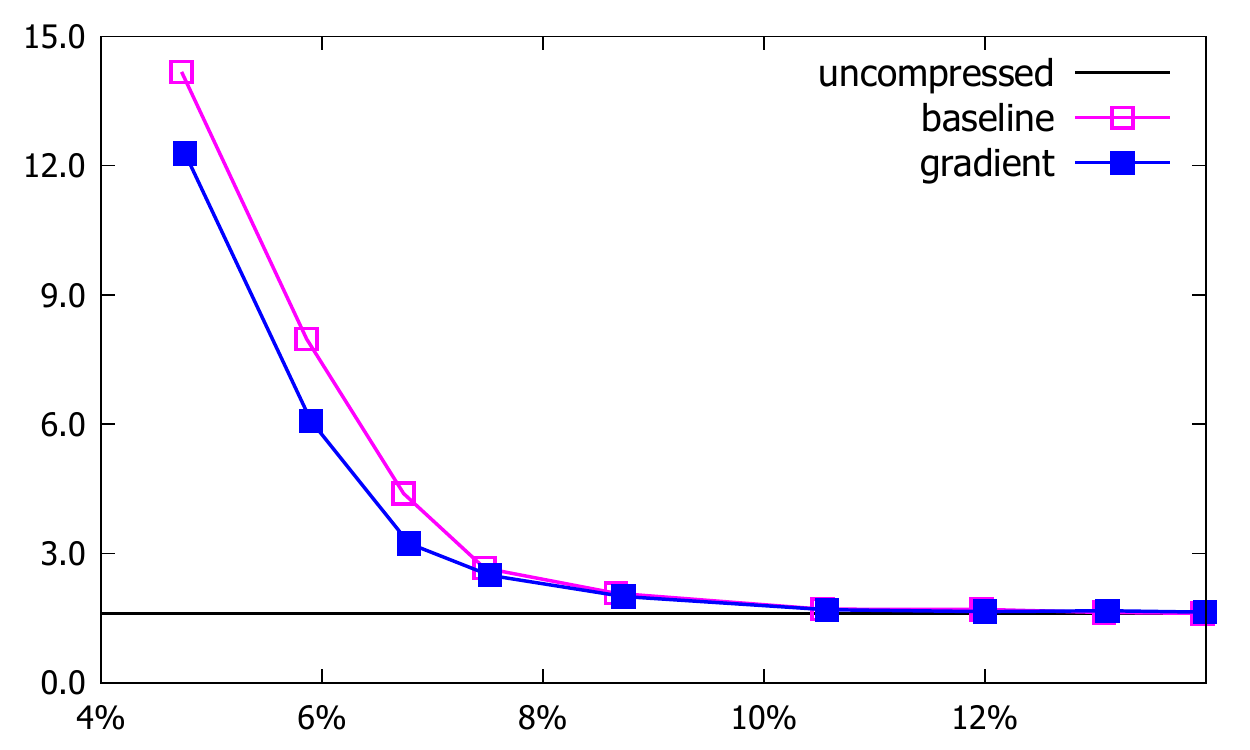}
    \put(-148,15){\rotatebox{90}{Cross Entropy}}
    \put(-110,-10){Compression Ratio}
    \put(-260,-10){Compression Ratio}
    \put(-400,-10){Compression Ratio}
    \caption{Result for supervised quantization experiment for CIFAR 100 experiment. Left: top-1 accuracy. Middle: top-5 accuracy. Right: cross entropy loss.}
    \label{fig:ce_quantization_100}
\end{figure}

In Figure~\ref{fig:ce_pruning_100} and Figure~\ref{fig:ce_quantization_100}, we show the result for supervised pruning and quantization, introduced in Section~\ref{sec:experiment}.2. Due to the slow running speed for estimating the Hessian $\nabla^2_{w_i} \mathcal{L}_w(x,y)$, we only compare two objectives --- baseline and gradient. It is shown that the gradient objective gives better accuracy and smaller loss.

\subsection{Algorithm for finding optimal quantization}

We present a variation of $k$-means algorithm which are used to find the optimal quantization for the following objective,
\begin{eqnarray*}
    \min_{c_1, \dots, c_k, A \in [k]^m} \sum_{i=1}^m \left( I_i (w_i - c_{A_i})^2 + H_i (w_i - c_{A_i})^4 \right)
\end{eqnarray*}
where $I_i$ is positive weight importance for quadratic term and $H_i$ is positive weight importance for quartic term. Basic idea of the algorithm is --- the assignment step finds the optimal assignment given fixed centroids, and the update step finds the optimal centroids given fixed assignments.
 This is used for gradient+hessian objective in Section~\ref{sec:experiment}.2. 
\begin{algorithm}[htbp]
    \caption{Quartic weighted $k$-means}
    \label{algo:quartic_weighted_k_means}
\begin{algorithmic}
    \INPUT Weights $\{w_1, \dots, w_m\}$, weight importances $\{I_1, \dots, I_m\}$, quartic weight importances $\{H_1, \dots, H_m\}$, number of clusters $k$, iterations $T$ \\
    \STATE {\bf Initialize} the centroid of $k$ clusters $\{c_1^{(0)}, \dots, c_k^{(0)}\}$
    \FOR{$t=1$ to $T$}
        \STATE {\bf Assignment step:}
            \FOR{$i=1$ to $m$}
                \STATE Assign $w_i$ to the nearest cluster centroid, i.e. $A_i^{(t)} = \arg\min_{j \in [k]} (w_i - c_j^{(t-1)})^2$.
            \ENDFOR
        \STATE {\bf Update step:}
            \FOR{$j=1$ to $k$}
                \STATE Find the only real root $x^*$ of the cubic equation
                \begin{eqnarray*}
                    (\sum_{i: A_i^{(t)} = j} 4H_i) x^3 - (\sum_{i: A_i^{(t)} = j} 12 H_i w_i) x^2 + (\sum_{i: A_i^{(t)} = j} (12 H_i w_i^2 + 2 I_i)) x - (\sum_{i: A_i^{(t)} = j} (4 H_i w_i^3 + 2 I_i w_i)) = 0
                \end{eqnarray*}
                \STATE Update the cluster centroids $c_j^{(t)}$ be the real root $x^*$.  
            \ENDFOR
    \ENDFOR
    \OUTPUT Centroids $\{c_1^{(T)}, \dots, c_k^{(T)}\}$ and assignments $A^{(T)} \in [k]^m$.
\end{algorithmic}
\end{algorithm}

Here we show that the cubic equation in Algorithm~\ref{algo:quartic_weighted_k_means} has only one real root. It was know that if the determinant $\Delta_0 = b^2 - 3ac$ of a cubic equation $a x^3 + b x^2 + cx + d = 0$ is negative, then the cubic equation is strictly increasing or decreasing, hence only have one real root. Now we show that the determinant is negative in this case (we drop the subsripts of the summation for simplicity).
\begin{eqnarray*}
\Delta_0 &=& (\sum 12 H_i w_i)^2 - 3 (\sum 4 H_i) (\sum 12 H_i w_i^2 + 2 I_i) \,\notag\\
&=& 144 \left( (\sum H_i w_i)^2 - (\sum H_i) (\sum H_i w_i^2) \right) - 24 (\sum H_i) (\sum I_i)
\end{eqnarray*}
The first term is non-positive because of Cauchy-Schwarz inequality. The second term is negative since $H_i$'s and $I_i$'s are all positive. Hence the determinant is negative.

\subsection{Effects of hyperparameters}
Here we briefly talk about the hyperparameters used in estimating the gradients $\mathbb{E} [\nabla_{w_i} \mathcal{L}_w(X,Y)]$ and hessians $\mathbb{E} [\nabla^2_{w_i} \mathcal{L}_w(X,Y)]$.

\subsubsection{Temperature scaling method}
The temperature scaling method proposed by~\cite{guo2017calibration}, aims to improve the confidence calibration of a classification model. Denote $z_w(x) \in \mathbb{R}^C$ is the output of the neural network, and classical softmax gives $f^{(c)} _w(x)= \frac{\exp\{z^{(c)}_w(x)\}}{\sum_{c \in C} \exp\{z^{(c)}_w(x)\}}$. The temperature sclaed softmax gives
\begin{eqnarray*}
f^{(c)}_w(x) = \frac{\exp\{z^{(c)}_w(x)/T\}}{\sum_{c \in C} \exp\{z^{(c)}_w(x)/T\}}
\end{eqnarray*}
by choosing different $T$, the prediction of the model does not change, but the cross entropy loss may change. Hence, we can finetune $T$ to get a better model calibration. In our experiment, we found that in MNIST experiment, the model is poorly calibrated. Hence, the variance of estimating gradient and hessian is very large. To solve this, we adopt a temperature $T > 1$ such that the loss from correctly-predicted data can also be backpropagated.

\begin{figure}[htbp]
    \centering
	\includegraphics[width=0.4\textwidth]{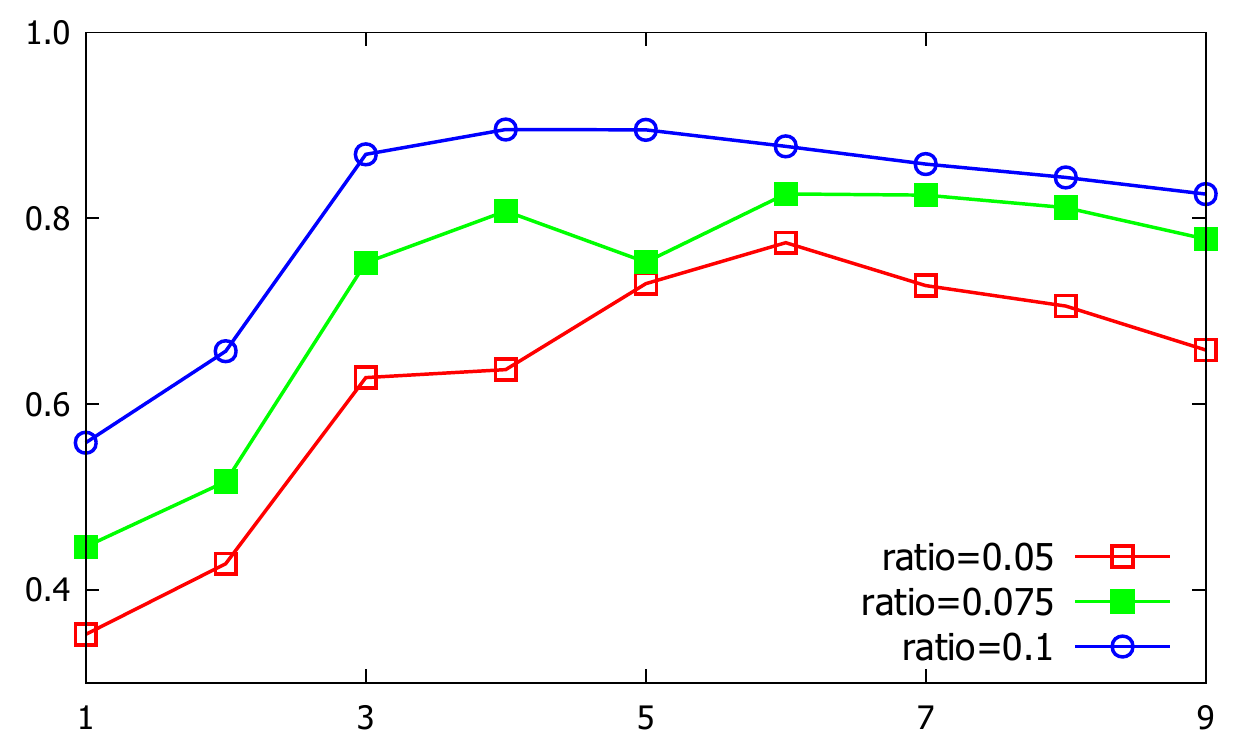}
	\put(-200,40){\rotatebox{90}{Accuracy}}
    \includegraphics[width=0.4\textwidth]{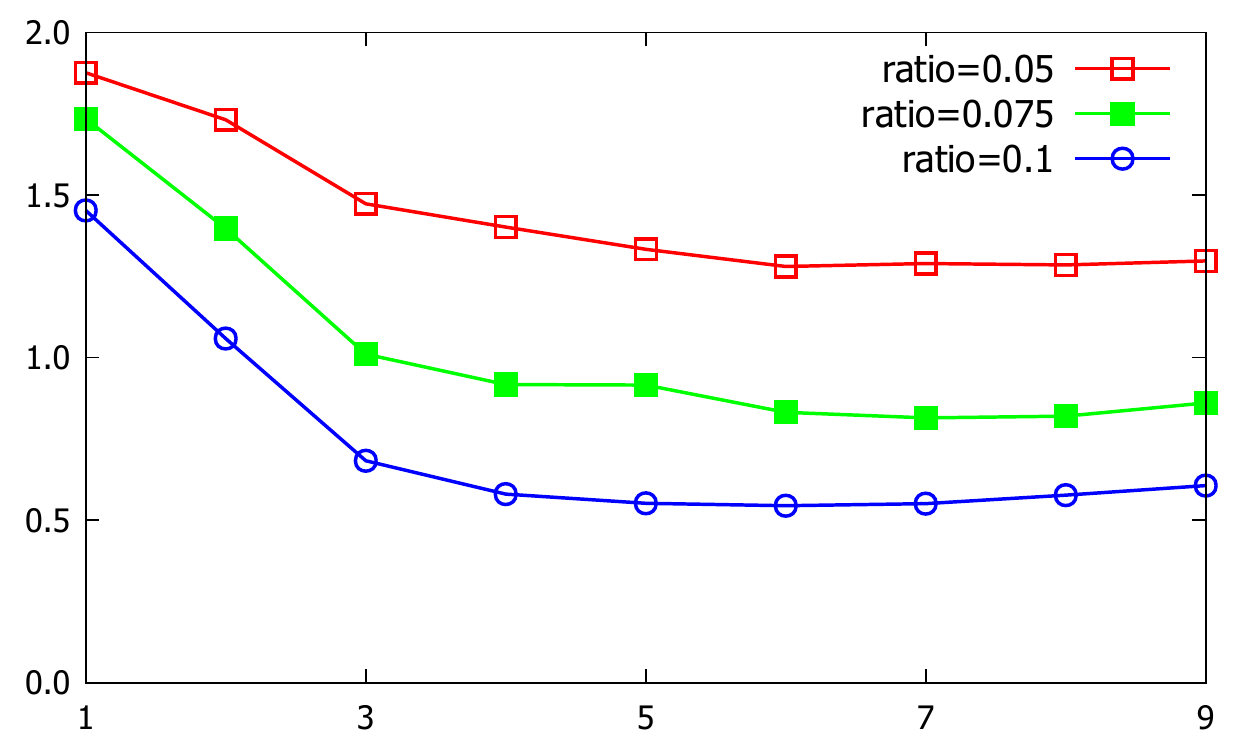}
    \hspace{0.5in}
    \put(-192,30){\rotatebox{90}{Cross Entropy}}
    \put(-130,-10){Temperature}
    \put(-320,-10){Temperature}
    \caption{Effect of the temperature $T$. Left: accuracy of supervised pruning for MNIST. Right: cross entropy of supervised pruning for MNIST. Different lines denote different compression ratio $\in \{0.05,0.075,0.1\}$}
    \label{fig:T}
\end{figure}

In Figure~\ref{fig:T}, we show the effect of $T$ for supervised pruning for MNIST. We can see that as $T$ increases from 1, the performance become better at first, then become worse. In our experiment, we choose $T \in \{1.0, 2.0, \dots, 9.0\}$ which gives best accuracy.

\subsubsection{Regularizer of hessian}
\begin{figure}[htbp]
    \centering
	\includegraphics[width=0.4\textwidth]{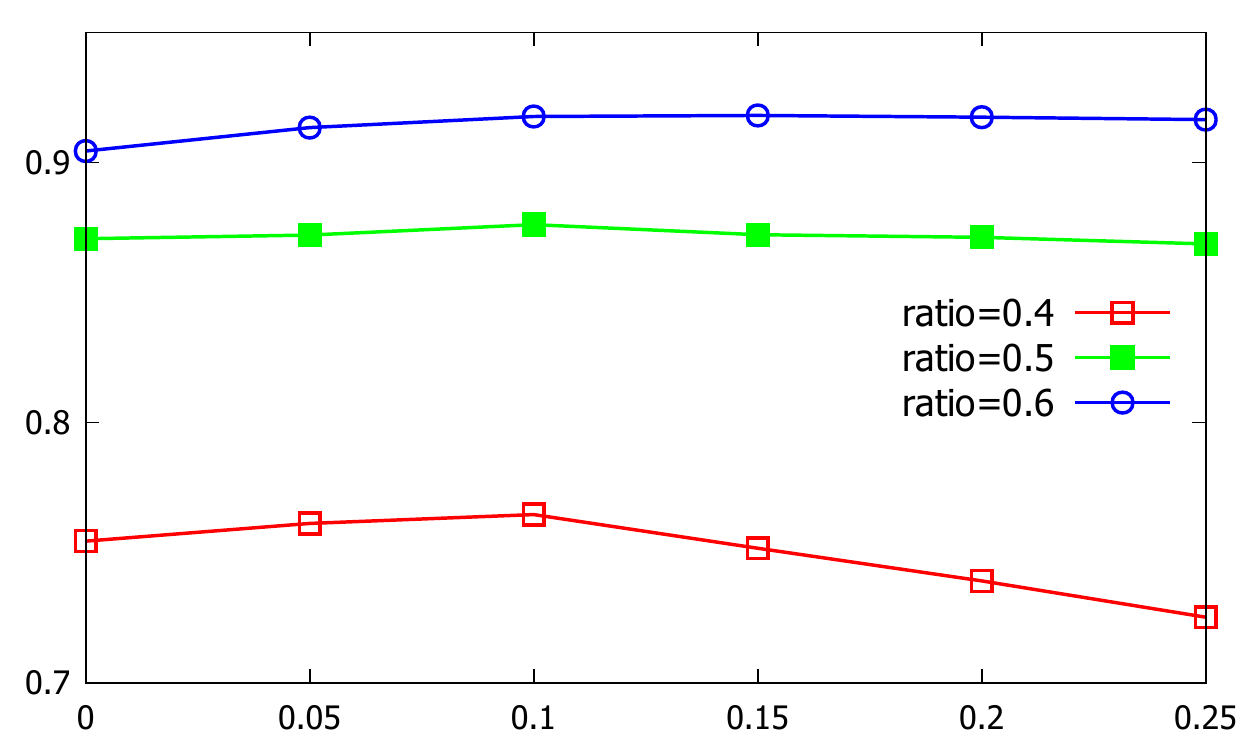}
	\put(-200,40){\rotatebox{90}{Accuracy}}
    \includegraphics[width=0.4\textwidth]{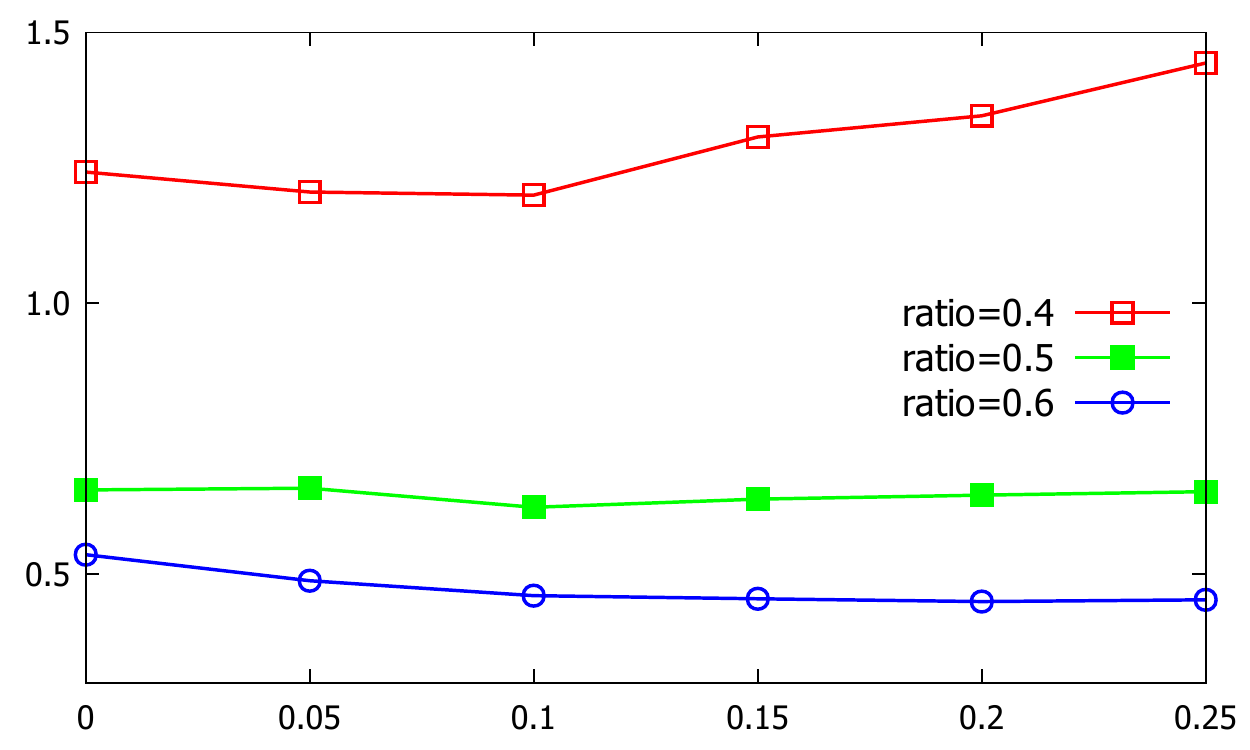}
    \hspace{0.5in}
    \put(-192,30){\rotatebox{90}{Cross Entropy}}
    \put(-100,-10){$\mu$}
    \put(-290,-10){$\mu$}
    \caption{Effect of the regularizer $\mu$. Left: accuracy of supervised pruning for CIFAR10. Right: cross entropy of supervised pruning for CIFAR10. Different lines denote different compression ratio $\in \{0.4,0.5,0.6\}$}
    \label{fig:mu}
\end{figure}

In the experiments, we estimate the hessians $\mathbb{E} [\nabla^2_{w_i} \mathcal{L}_w(X,Y)]$ using the curvature propagation algorithm~\cite{martens2012estimating}. However, due to the sparsity introduced by ReLU, there are many zero entries of the estimated hessians, which hurts the performance of the algorithm. Hence, we add a constant $\mu > 0$ to the estimated hessians. In Figure~\ref{fig:mu}, we show that effect of $\mu$ for supervised pruning for CIFAR10. We can see that as $\mu$ increases from 0, the performance increase first then decrease. We use simple binary search to find the best $\mu$.

\end{document}